\documentclass[1p, final]{elsarticle}

\usepackage{subfig}
\usepackage{amsmath,bm}
\usepackage{color}
\usepackage{enumerate}
\usepackage{amssymb}
\usepackage{epstopdf}
\usepackage{epsfig}
\usepackage[table,xcdraw]{xcolor}
\usepackage{algpseudocode,algorithm,algorithmicx}

\usepackage{bigstrut}
\usepackage{lineno,hyperref}
\modulolinenumbers[5]

\journal{Applied Mathematical Modelling}

\begin{document}

\begin{frontmatter}

\title{Localized Nonlinear Solution Strategies for Efficient Simulation of Unconventional Reservoirs}

\author{Jiamin Jiang}
\cortext[mycorrespondingauthor]{Corresponding author}
\ead{jiamin.jiang@chevron.com}

\address{Chevron Energy Technology Co.}
\address{1500 Louisiana St., Houston, TX 77002, USA}

\date{July 29, 2020}

\begin{abstract}

Accurate and efficient numerical simulation of unconventional reservoirs is challenging.~Long periods of transient flow and steep potential gradients occur due to the extreme conductivity contrast between matrix and fracture.~Detailed near-well/near-fracture models are necessary to provide sufficient resolution, but they are computationally impractical for field cases with multiple hydraulic-fracture stages. 

Previous works in the literature of unconventional simulations mainly focus on the gridding level that adapts to wells and fractures. Limited research has been conducted on nonlinear strategies that exploit locality across timesteps and nonlinear iterations. It was reported that an individual Newton update is typically sparse and nonlinear convergence is constrained by a small portion of the model. To perform localized computations, an $\textit{a-priori}$ strategy is essential to first determine the active subset of simulation cells for the subsequent iteration.~The active set flags the cells that will be updated, and then the corresponding localized linear system is solved. 

The objective of this work is to develop localization methods that are readily applicable to complex fracture networks and flow physics in unconventional reservoirs.~By utilizing the diffusive nature of pressure updates, an adaptive algorithm is proposed to make adequate estimates for the active domains.~In addition, we further develop a localized solver based on nonlinear domain decomposition (DD). Compared to a standard DD method, domain partitions are dynamically constructed.~The new solver provides effective partitioning that adapts to flow dynamics and Newton updates. 

We evaluate the developed methods using several complex problems with discrete fracture networks.~The problems consider multi-phase and compositional fluid systems with phase changes.~The results show that large degrees of solution locality present across timesteps and iterations. Compared to a standard Newton solver, the new solvers enable superior computational performance.~Moreover, Newton convergence behavior is preserved, without any impact on solution accuracy.

\end{abstract}

\end{frontmatter}

\section{Introduction}

Unconventional reservoirs have received great attention as a primary energy resource in the past decade worldwide. Economic production from these reservoirs depends on effective stimulation by means of hydraulic fracturing. Micro-seismic measurements and other evidence suggest the creation of complex fracture networks that connect huge reservoir surface areas to the wellbore [1-3].~In terms of reservoir development and management, numerical simulation continues to play a critical role in evaluating and optimizing the stimulation and production processes [4,5].

Because of the ultra-low permeability of matrix, a long period of transient flow occurs in unconventional formation. The extreme contrast in conductivity between matrix and fracture also results in steep potential gradients that are difficult to capture. Therefore, detailed near-well/near-fracture models are necessary to provide sufficient resolution for the matrix-fracture interactions [3,4,6].~However, a fine grid simulation requires too much CPU time and it is impractical to perform for the entire domain in field cases with multiple hydraulic-fracture stages [5,7,8].

Several modeling approaches for fractured-well have been proposed in the literature, attempting to improve the solution efficiency while maintaining the accuracy. One simple approach is applying local grid refinement (LGR) on a coarse background grid [4,8-14]. A number of meshing algorithms are available to generate adaptive and optimized mesh with good quality around discrete fractures. Pruess and Narasimhan [15] developed a sub-gridding method called multiple interacting continua (MINC) by subdividing each matrix cell according to the distance from fractures. The MINC method is intended to improve the classical dual-porosity model [16,17], and subsequently better characterize the transient effects with sharp solution gradients. In recent years, MINC was widely applied and extended for the simulations of unconventional reservoirs [12, 18-23]. In addition, Ding et al. [7] proposed a coupled modeling method that combines a coarse-grid reservoir model with detailed near-fracture models. The two models are solved, and the associated boundary conditions are updated in an alternate mode. The coupled method can be viewed as dynamic upscaling that computes the time-dependent fracture index for the coarse domain. Although some promising results were presented, the detailed models may still take a large fraction of the overall computational expense. 

In reservoir simulation, the Fully Implicit Method (FIM) is often used for the temporal discretization of the conservation equations [24,25]. FIM offers unconditional stability, but it requires the solution of large coupled nonlinear systems. For a target timestep, a sequence of Newton iterations is performed until convergence. This iterative process is expensive and can account for a significant fraction of the total cost. Here define $n$ as the total number of degrees of freedom in a system.~Considering the costs of computing the residual vector, Jacobian matrix, and thermodynamic properties, the overall complexity of a nonlinear iteration is generally superlinear in $n$ [26]. Recently, an algebraic dynamic multilevel method (ADM) was introduced for fully implicit simulations of flow and transport in heterogeneous porous/fractured media [13,27,28]. Built on the fine-scale FIM discrete system, ADM constructs a multilevel FIM system on a dynamic nested grid. The test results showed that ADM provided accurate solutions by employing only a fraction of the fine-scale grid-cells.

Previous works in the literature of unconventional simulations mainly focus on the gridding level that adapts to wells and fractures. Limited research has been conducted on solver techniques that exploit locality across timesteps and nonlinear iterations. The benefit of this type of methods is evident: the solution accuracy is maintained, because neither discretization scheme nor spatial mesh will be modified [26,29]. Simulation studies have shown that flow dynamics evolve quite slowly within the ultra-tight formation. Pressure drop may remain in the vicinity of fractures even after years of production [4,7,20]. Conceivably, a significant speedup can be achieved if performing adaptive computations only for the locales that are undergoing changes. In addition to the timestep level, a large degree of locality also presents on the nonlinear (Newton) level. It was reported that an individual Newton update is typically sparse and the nonlinear convergence is constrained by a small portion of the model [26,29,30]. To exploit the locality at each iteration, an $\textit{a-priori}$ strategy is essential to first identify the active subset of simulation cells. The active set flags the cells that will be updated, and then the corresponding localized linear system is solved. Lu and Beckner [30] observed that over the course of several iterations, the sparsity pattern of the Newton updates was related to that of the discrete residual vector. They proposed to use non-zero entries in the residual vector as an estimate of the active set for the subsequent iteration. It should be mentioned that their heuristic strategy may suffer from an efficiency issue due to overly conservative estimate. Sheth and Younis [29] have shown that missing any non-zero update during the localization process may lead to worse nonlinear convergence, compared to the standard Newton method. A theoretical framework was then developed to predict the sparsity pattern of Newton updates. Analytical derivations were made to ensure a conservative estimate of the active sets. The results in Sheth and Younis [29] demonstrated that their localization method performs quite well for several challenging models. 

In this work we do not intend to rely on an analytical derivation or conservative estimate for the active sets. The objective is to develop localization methods that readily accommodate to complex fracture networks and flow physics for the simulations of unconventional reservoirs. Through aggressive localization, the computational speedup is expected to be greatly improved. We recently revealed that the Newton updates for pressure-driven problems exhibit diffusive and global behaviors, because of its parabolic nature [31]. By utilizing the nature of pressure updates, an adaptive algorithm is proposed to make adequate estimates for the active domains. In addition, we further develop a localized solver based on nonlinear domain decomposition (DD). Compared to a standard DD method, domain partitions are dynamically constructed from the previous iterations. During the nonlinear DD process, the subproblems of an iteration are solved sequentially, and thus the localization can be naturally achieved. This leads to a reliable strategy that exploits the locality while preserving the convergence behavior of the standard Newton process. Note that the two methods developed involve different complexities and efforts for implementation. Subsequently, their applications depend on specific efficiency and implementation considerations.

We evaluate the localization methods using several complex problems with discrete fracture networks. The test problems consider multi-phase and compositional fluid systems with phase changes. The results show that large degrees of solution locality present across timesteps and iterations. Compared to a standard Newton solver, the new solvers exhibit superior computational performance. Moreover, Newton convergence behavior is preserved, without any impact on the solution accuracy.

\section{Isothermal compositional model}


We consider compressible gas-oil flow in porous media without capillarity. We ignore water that does not exchange mass with the hydrocarbon phases. 

The conservation equations for the isothermal compositional problem containing $n_c$ components are written as, 
\begin{equation} 
\label{eq:mass_con_comp}
\frac{\partial}{\partial t } \left [ \phi \left ( x_c \rho_{o} s_{o} + y_c \rho_{g} s_{g} \right ) \right ] + \nabla \cdot \left ( x_c \rho_{o} \textbf{u}_{o} + y_c \rho_{g} \textbf{u}_{g} \right ) - q_{c} = 0,
\end{equation}
where $c \in \left \{ 1,...,n_c \right \}$. $x_{c}$ and $y_{c}$ are molar fractions of component $c$ in the oil and gas phases, respectively. $\phi$ is rock porosity and $t$ is time. $\rho_l$ is phase molar density. $s_{l}$ is phase saturation. $q_{c}$ is well flow rate. 

Phase velocity $\textbf{u}_l$ is expressed as a function of phase potential gradient $\nabla \Phi_l $ using the extended Darcy's law,
\begin{equation} 
\label{eq:phase_vel}
\textbf{u}_l = - k \lambda_l \nabla \Phi_l = -k\lambda_l\left ( \nabla p - \rho_l g \nabla h \right ).
\end{equation}
where $k$ is rock permeability. $p$ is pressure. Capillarity is assumed to be negligible. $g$ is gravitational acceleration and $h$ is height. Phase mobility is given as $\lambda_{l} = k_{rl}/\mu_l$. $k_{rl}$ and $\mu_l$ are relative permeability and viscosity, respectively. 

In order to close the nonlinear system, additional equations are needed. These include the thermodynamic equilibrium constraints,
\begin{equation} 
\label{eq:vle_fu}
f_{c,o}(p,\textbf{x}) - f_{c,g}(p,\textbf{y}) = 0 ,
\end{equation}
where $p$, $T$, and $z_c$ denote pressure, temperature, and overall molar fraction, respectively. $f_{c,l}$ is the fugacity of component $c$ in phase $l$. 

We now write the phase constraints, 
\begin{equation} 
\label{eq:phase_const}
\sum_{c=1}^{n_c} x_{c} - 1 = 0, \qquad \sum_{c=1}^{n_c} y_{c} - 1 = 0,
\end{equation}
and the saturation constraint as,
\begin{equation} 
\label{eq:satu_const}
s_o + s_g - 1 = 0.
\end{equation}

The above system of equations provide a complete mathematical statement for two-phase multi-component flow. The local equilibrium constraints are enforced only when both phases are present.

\section{Natural-variables formulation}

An important aspect of any compositional formulation is the choice of dividing the equations and unknowns into primary and secondary sets. Here we employ the popular natural-variables set [25,32]. The primary unknowns include pressure, saturations, and molar fractions, 

(1) $p$ $-$ pressure [1],

(2) $s_l$ $-$ phase saturations [2], 

(3) $x_{c}$, $y_{c}$ $-$ phase compositions of each component [2$n_c$]. 
\\
The size of each variable is given in square bracket.

The various coefficients can be obtained as functions of the base variables. For a two-phase cell, the molar phase fraction is related to saturation as follows, 
\begin{equation} 
\nu_l = \frac{\rho_l s_l}{\sum_{m} \rho_m s_m}
\end{equation}
and overall molar fraction of component $c$ is written as, 
\begin{equation} 
z_c = x_{c} \nu_o + y_{c} \nu_g 
\end{equation}

Note that for single-phase ($l$) mixture, $\nu_l = s_l = 1$, and $x_{c,l} \equiv z_c$.

\subsection{Variable substitution}


An essential ingredient of the natural-variables formulation is the `variable substitution' process [32]. A common strategy for variable-switching between Newton iterations during a timestep is,

1. For any cell whose status in the previous iteration is single-phase, run the phase stability test [33] to check if the mixture becomes two-phase. For the mixture that splits into two phases, perform the flash to compute the phase compositions [34].

2. If a cell is already in the two-phase state, the thermodynamic constraints are included in the nonlinear system as part of the global Jacobian.

3. If a phase saturation, or phase fraction, becomes negative between two successive iterations, the phase disappears, and appropriate variable-switching is performed. 

The system of conservation equations is solved for single-phase regimes, and the combination of conservation equations and thermodynamic constraints is solved for the two-phase regime.

\subsection{Phase behavior}

Phase behavior computation is usually a stand-alone procedure for detecting phase changes. For a mixture of $n_c$ components and two phases, the mathematical model describing the thermodynamic equilibrium is [32], 

\begin{equation} 
\label{eq:vle_fu}
f_{c,o}(p,\textbf{x}) - f_{c,g}(p,\textbf{y}) = 0 ,
\end{equation}

\begin{equation} 
\label{eq:mass_zc}
z_c - \nu_o x_{c} - \left( 1 - \nu_o \right) y_{c} = 0 ,
\end{equation}

\begin{equation} 
\label{eq:x_y_equ}
\sum_{c=1}^{n_c} \left ( x_{c} - y_{c} \right ) = 0.
\end{equation}
where $\nu_l$ is molar fraction of phase $l$. We assume that $p$, $T$, and $z_c$ are known. The objective is to find all the $x_{c}$, $y_{c}$ and $\nu_l$. 

Phase behavior of a hydrocarbon mixture is commonly described using an Equation of State (EoS) model.

\section{Nonlinear solution strategies}

The spatial and temporal discretization schemes used for the compositional flow model are summarized in Appendix A.

\subsection{Newton method}

At each timestep of a FIM simulation, given the current state $u^{n_t}$, and a fixed timestep size $\Delta t$, we seek to obtain the new state $u^{n_t+1}$. 

The nonlinear residual system is solved by the Newton method,
\begin{equation} 
F(u^{n_t+1}) = 0 
\end{equation}

The Newton method generates a sequence of iterates, $u^{\nu}$, $\nu=0,1,...$, each involving the construction of a Jacobian matrix and solution of the resulting linear system, 
\begin{equation} 
J(u^{\nu }) \: \delta u^{\nu} = - F(u^{\nu })
\end{equation}
where
\begin{equation} 
\delta u^{\nu} = u^{\nu +1} - u^{\nu }
\end{equation}
and $J(u) = \frac{\partial F}{\partial u} (u)$ denotes the Jacobian matrix of $F$ with respect to $u$. 

Here we assume that entries of Newton update such that $\left | \delta u_i \right | < \epsilon$ are essentially negligible. Given a Newton iteration, the support set is defined for the indices of cells that exhibit non-zero update, 
\begin{equation} 
\textrm{supp} \, \delta u = \left \{ i \! : \left | \delta u_i \right | \geq \epsilon \, , \ \, i=1,...,n \right \}
\end{equation}

\subsection{Locality within solution processes}

In this work we focus on the solution process for the pressure-driven production problem with multi-phase multi-component fluid. We aim to exploit two levels of locality for improving computational efficiency. The first is on the timestep level. Because of the ultra-low matrix permeability in unconventional formation, transient flow within matrix may last a long period. As a result, flow dynamics (e.g. pressure propagation) evolve slowly and locally. During early stage of production, only a small portion of domain undergoes considerable variable changes. 

For a timestep, the solution update is the sum of all the corresponding Newton updates. Conceivably, the locality also presents on the nonlinear (Newton) level, even if most of the domain is affected over the timestep. Previous works showed that an individual update computed for flow and transport problem is typically sparse and constrained by a small subset of cells [26,29,30].

\begin{figure}[!htb]
\centering
\subfloat[Iteration 1]{
\includegraphics[scale=0.41]{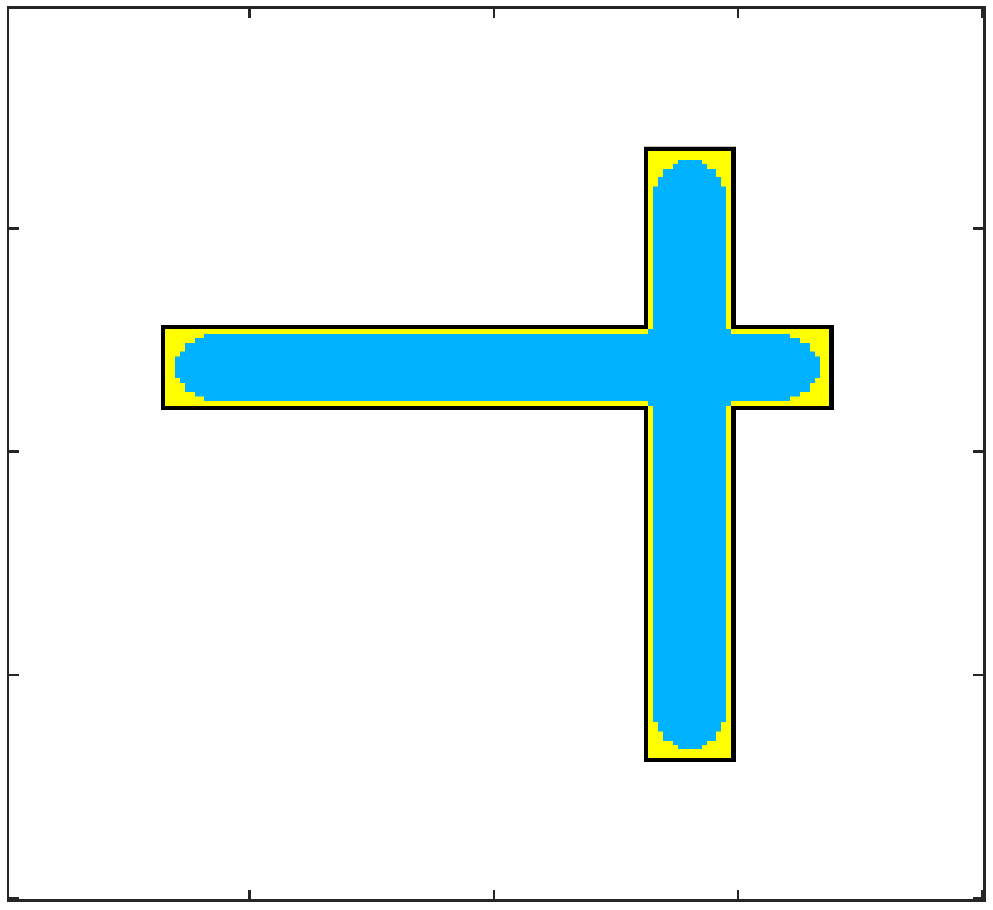}}
\
\subfloat[Iteration 2]{
\includegraphics[scale=0.41]{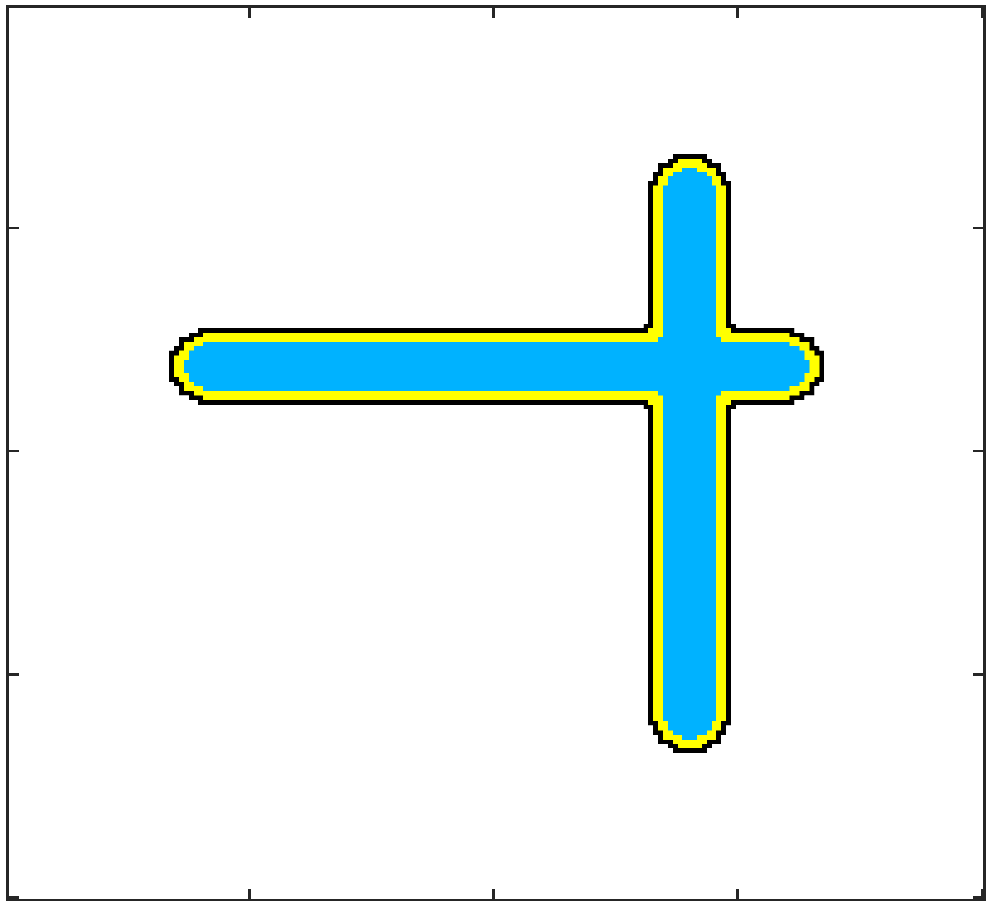}}
\
\subfloat[Iteration 3]{
\includegraphics[scale=0.41]{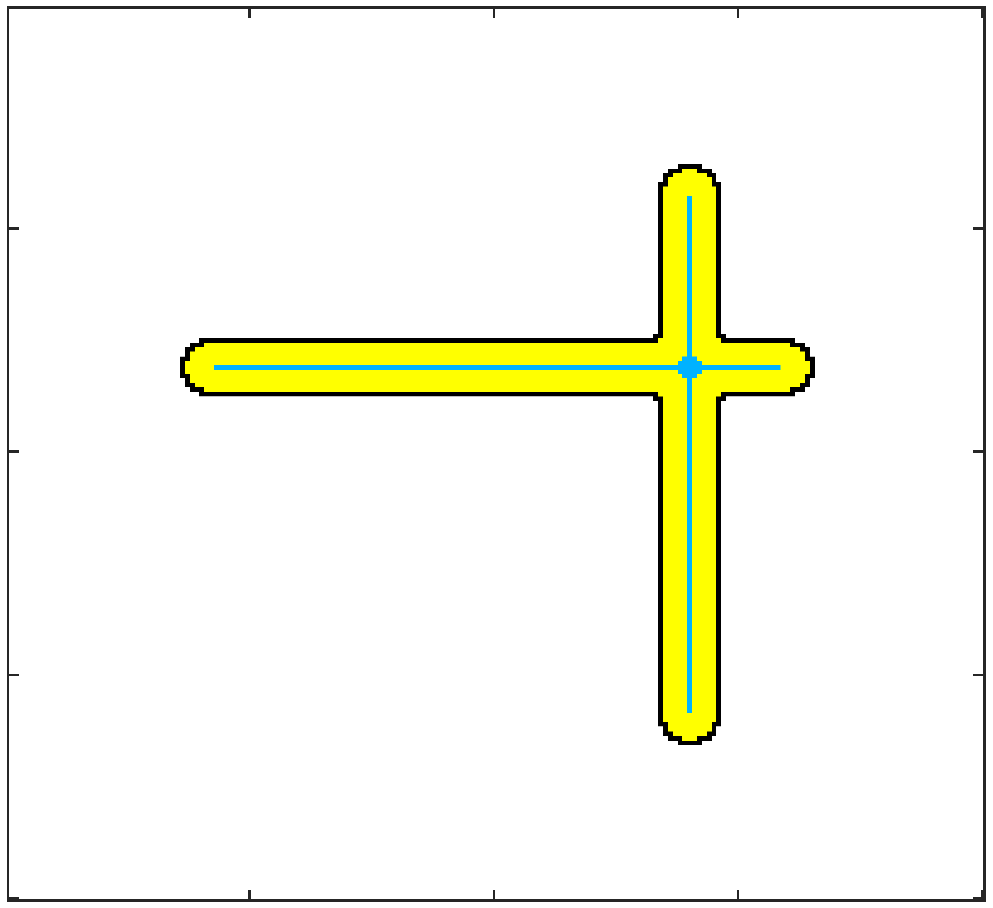}}
\caption{Flagging of the pressure updates for the three Newton iterations of a timestep.}
\label{fig:flag_2frac}
\end{figure}

Here we show an example with two-fracture to demonstrate the solution behavior. The details of the model will be given in the result section.~The flagging profiles for the Newton iterations of a timestep are plotted in \textbf{Fig.~\ref{fig:flag_2frac}}. The cells that exhibit non-zero pressure updates are flagged in color blue.~As we can see, the first iteration reaches the maximum area, and the region gets smaller as the iterations proceed. In the last iteration, the support set of the updates localizes to just a few cells.

Lu and Beckner [30] proposed an adaptive Newton strategy that solves localized systems.~Their method identifies unconverged cells and their neighbors as the active subset to be updated. The unconverged set is inflated to a heuristic extent (e.g. plus one layer of neighbor cells), as a safety measure. 

From the simulation studies, we also observe that under a convergent Newton sequence, the support set of updates always shrinks after each iteration, such that, 
\begin{equation} 
\textrm{supp} \, \delta u^{\nu +1} \subseteq \textrm{supp} \, \delta u^{\nu}
\end{equation}

Therefore, the support set from a current Newton iteration becomes an adequate estimate for the subsequent iteration.

The adaptive algorithm by Lu and Beckner [30] is simple to implement. However, the algorithm starts with the entire domain at the first iteration of a timestep, to ensure conservative estimates for the active set. This will considerably contribute to the overall computational cost. 

Our recent studies revealed that Newton iterations are closely tied to the underlying physics problem [31]. The updates for a hyperbolic transport problem may have local support that propagates through the domain as the Newton process goes forward. By comparison, the support of a pressure (flow) problem shows diffusive and global behaviors, due to its parabolic nature. To exploit this mechanism, here we propose a localized Newton strategy, which is aggressive in the way that it does not require an initial conservative estimate for the active domain. We observe that the support of pressure updates tends to reach the solution front at the first iteration.~Therefore, the localized algorithm can start with a moderate active domain $\Omega_A$, and expand it if necessary using the outermost (boundary) layer of $\Omega_A$.

\subsection{Localized Newton algorithm}

\begin{figure}[!htb]
\centering
\includegraphics[scale=0.45]{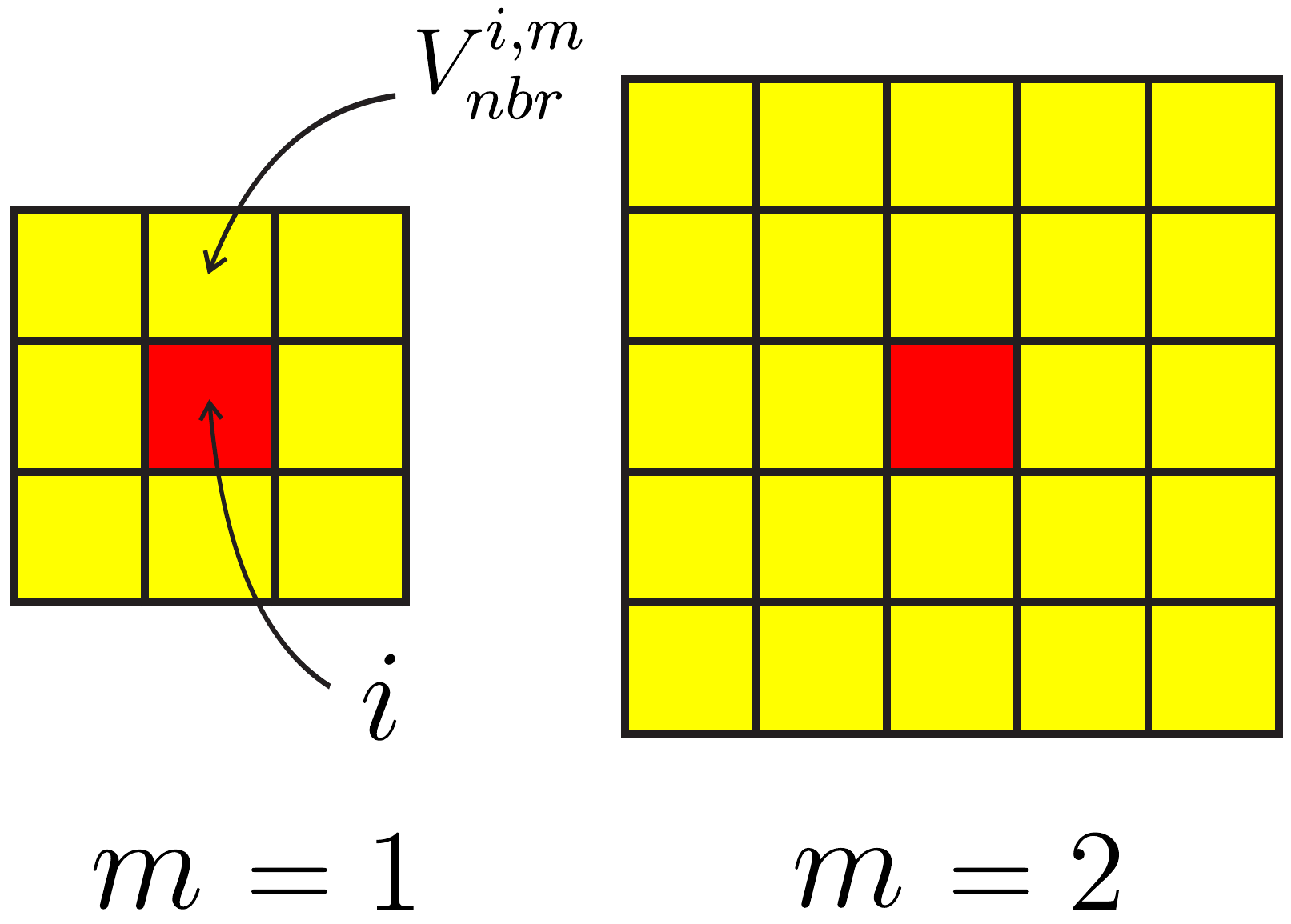}
\caption{Illustration for the neighbors of cell $i$.}
\label{fig:neigh_local}
\end{figure}

We describe the algorithm based on the proposed localized Newton strategy. Consider a nonlinear system of equations $F = \left ( F_1 , ... , F_n \right )^T$ with unknowns $u = \left ( u_1, ... , u_n \right )^T$. Let $V = \left ( 1, ... , n \right )$ be an index set; i.e., there is one integer for each $F_i$ and unknown $u_i$. Let $V_A$ be the index set that contains the active cells, and $n_A$ be the dimension of $V_A$. We define $V_B \subseteq V_A$ as the cell set for the outermost (boundary) layer of the active domain. Further define the set $V_{nbr}^{i, m}$ for the neighbors of cell $i$, and $m$ is the number of layers that are incorporated. The illustration for $V_{nbr}^{i, m}$ with $m=1$ and $m=2$ is plotted in \textbf{Fig.~\ref{fig:neigh_local}}. The neighbor cells are flagged in yellow.

Let $R_A$ be a Boolean matrix of dimension $n_A \times n$. $R_A$ corresponds to the restriction operator from $V$ to $V_A$. The transpose matrix $R_A^T$ is an extension operator from $V_A$ to $V$. Then the nonlinear function and the local unknowns of the active set $V_A$ can be expressed as $F_A = R_A \, F$ and $u_A = R_A \, u$, respectively. The Jacobian of $F_A$ is, 
\begin{equation} 
J_A = R_A \, J \, R_A^T 
\end{equation}
Note that the submatrix $J_A$ can be directly constructed, and thus the full matrix $J$ is never required. There is no need to evaluate the cell properties, residual and Jacobian for the inactive domain.

The localization method takes as input a state at time level $n_t$ and outputs the updated state. The algorithmic details can be described as, 

\begin{enumerate}[\quad Step 1.]
\item Set the iteration counter $\nu$ to zero, and initialize $u^{\nu}$ to the current state $u^{n_t}$. Construct the initial active set $V_A$ and the associated boundary set $V_B$. 
\item Perform localization: solve the reduced linear system over the active domain, and update the solution. 
\item For each cell $i$ in $V_B$, if the update is larger than the cutoff value, inflate $V_A$ with the neighbor subset $V_{nbr}^{i, m}$. 
\item If there is any non-negligible update in $V_B$, the localization status stays in the `expand' mode; otherwise, switch to the `shrink' mode, and $V_A$ is specified as the support set of $\delta u_A$.
\item Check convergence criteria of Newton iterations. If not converged, go back to Step 2. Otherwise, the timestep is finished.
\end{enumerate}

\begin{algorithm}
\caption{Localized Newton} \label{alg:local_new}
\smallskip
\begin{algorithmic}[1]
\State $\nu = 0, \ u^{\nu} = u^{n_t}$
\State Initialize $V_A$ and $V_B$.
\smallskip
\While{ \textit{is\_expand} $\, \textrm{or} \,$ \textit{not\_converged} }
\Comment{Newton loop}
\State Solve reduced linear system,

$J_A \, \delta u_A = - F_A$
\smallskip
\State $u^{\nu + 1} = u^{\nu} + R_A^T \, \delta u_A$
\smallskip
\State Check convergence criteria.
\smallskip
\For{$i \in V_B$}
\Comment{Determine active set}
\If{$\left | \delta u_{i,A} \right | \geq \epsilon$}
\State Obtain $V_{nbr}^{i, m}$ of cell $i$,
\smallskip
\State $V_A = V_A \cup V_{nbr}^{i, m}$
\EndIf
\EndFor
\smallskip

\If{$\left \| \delta u_{B} \right \|_{\infty} < \epsilon$}
\Comment{Shrink mode}
\State $\textit{is\_expand} \, \leftarrow \textit{false}$
\State $V_A = \textrm{supp} \, \delta u_A$
\Else
\State $\textit{is\_expand} \, \leftarrow \textit{true}$
\EndIf
\smallskip
\State Update $V_B$ from $V_A$.
\smallskip
\State $\nu \leftarrow \nu + 1$
\smallskip
\EndWhile 
\end{algorithmic}
\end{algorithm}

The proposed method is also outlined in Algorithm \ref{alg:local_new}. We can see that the localization process determines the active set that needs to be solved for the subsequent iteration. The process consists of two stages. In the first stage, $V_A$ does not fully cover the actual support of the timestep. Based on the diffusive nature of pressure updates, $V_B$ is used to detect and expand the active domain. The localized algorithm is self-adaptive in a sense that Newton update already provides an adequate estimate. 

The initial active set can be constructed to comprise the cells in the vicinity of wells or fractures.~Subsequently, computational speedup will be greatly increased.~On the other hand, the iterations taken during the `expand' mode may degrade the nonlinear convergence, compared to the standard Newton method. In practice, a balance needs to be achieved between more aggressive localization and increased number of iterations. 

It should be noted that the iterative process for expanding active domain can be viewed as a nonlinear domain decomposition (DD) problem. Each subproblem is solved with Dirichlet boundary (constant pressure) conditions.~Similarly to nonlinear preconditioning, the subdomain solutions can provide better initial guesses for Newton iterations. As a result, the localized solver shows satisfying convergence performance from the simulation cases.

\subsection{Adaptive Nonlinear Domain Decomposition}

For a localized Newton algorithm, more iterations may be required, if the estimate for the support set is not conservative. On the other hand, an excessively conservative estimate will reduce the speedup gained from the localized computations. 

In this work we further develop an adaptive method that provides aggressive localization, while preserving the convergence behavior of the standard Newton process. To present the method, consider first a nonlinear domain decomposition (DD) with non-overlapping partitions, 
\begin{equation} 
\bigcup_{k=1}^{N} V_k = V , \quad V_j \cap V_k = \varnothing \ \ \textrm{if} \ j \neq k , \quad \textrm{and} \ V_k \subset V.
\end{equation}

Let $n$ be the total number of unknowns and $n_k$ be the total number of unknowns associated with the subset $V_k$. The restrictions of $u$ and $F$ to $V_k$ are $u_{k} = R_{k} \, u$ and $F_{k} = R_{k} \, F$, respectively. The Jacobian of the subproblem $k$ is given as, 
\begin{equation} 
J_{k} = R_k \, J \, R_k^T 
\end{equation}
with $k = 1 , ..., N$. The boundary conditions for a subproblem are Dirichlet-type and taken from the neighboring subdomains. 

A global solution of the nonlinear DD method is obtained by solving first subproblems and then gluing them together [35],
\begin{equation} 
u^{\nu+1} = \sum_{k=1}^{N} R_k^T \, u_{k}^{\nu+1}
\end{equation}
It is common to apply the additive (Jacobi) form of the Schwarz methods. For a standard DD, static partitions of simulation grid are performed in a preprocessing step [35-37].

To exploit the localized behaviors of flow problems, we propose instead an adaptive DD solver based on dynamic partitions. Utilizing the diffusive nature of pressure updates, subdomains are constructed from the previous iterations. To achieve localized computations, the subproblems of a nonlinear DD iteration are solved sequentially. This leads to a multiplicative (Gauss-Seidel) Schwarz method. 

The algorithmic details of the adaptive DD method for a timestep are given as, 

\begin{enumerate}[\quad Step 1.]
\item Set the iteration counters $\nu$ and $k$ to zero, and initialize $u^{\nu}$ to the current state $u^{n_t}$. 
Construct the initial active set $V_A^k$, and the associated boundary sets $V_B^k$ and $V_{\partial B}^{k}$.
Define $\partial B$ as the outer layer adjacent to $\Omega_A$, such that $V_{\partial B}^{k} \cap V_A^k = \varnothing$. 

\item Start expanding the active domain, and perform localized computations. Construct the subdomains, and the associated cell sets, using the Newton updates. Specifically, for each cell $i$ in $V_B^k$, if the update is larger than the cutoff value, inflate $V_A^{k+1}$ and the total active set $V_T$ with the neighbor subset $V_{nbr}^{i, m}$. 

\item If there is any non-negligible update in $V_B^k$, remain in the `expand' mode, and obtain the boundary sets of $V_A^{k+1}$. Otherwise, switch to the `shrink' mode, indicating that the maximum support set over the timestep is reached.

\item Set $N$ as the number of the constructed subdomains. Start the nonlinear DD loop, with the counter $\nu$ denoting the outer iteration. For each $k$, first collect $\delta u_{\partial B}^{k}$ over $V_{\partial B}^{k}$, from the latest Newton updates.

If there is any non-negligible element in $\delta u_A^{k, \, \nu}$ or $\delta u_{\partial B}^{k}$, perform localized computation and update the solution.

\item Check convergence criteria. Repeat the outer iteration until all the subdomains are converged.

\end{enumerate}

\begin{algorithm}
\caption{Adaptive Nonlinear Domain Decomposition} \label{alg:local_nDD}
\smallskip
\begin{algorithmic}[1]
\State $\nu = 0 , \ k = 0$ 
\State $u^{\nu} = u^{n_t}$
\State Initialize $V_A^k$, $V_B^k$ and $V_{\partial B}^{k}$.
\State $V_T = V_A^k$
\smallskip
\While{ \textit{is\_expand} }
\Comment{Newton loop}
\State Local solve over $V_A^k$,

$J_A \, \delta u_A = - F_A$
\smallskip
\State $u^{\nu + 1} = u^{\nu} + R_A^T \, \delta u_A$

\medskip

\State $V_A^{k+1} \leftarrow \left \{  \right \}$
\smallskip
\For{$i \in V_B^k$}
\Comment{Determine active set}
\If{$\left | \delta u_{i,A} \right | \geq \epsilon$}
\State Obtain $V_{nbr}^{i, m}$ of cell $i$,
\smallskip
\State $V_T = V_T \cup V_{nbr}^{i, m}$
\smallskip
\State $V_A^{k+1} = V_A^{k+1} \cup V_{nbr}^{i, m}$
\smallskip
\EndIf
\EndFor

\smallskip

\If{$\left \| \delta u_{B} \right \|_{\infty} < \epsilon$}
\Comment{Shrink mode}
\State $\textit{is\_expand} \, \leftarrow \textit{false}$
\Else
\State $\textit{is\_expand} \, \leftarrow \textit{true}$
\smallskip
\State Obtain $V_{B}^T$ of $V_T$,
\smallskip
\State $V_B^{k+1} = V_{B}^T \cap V_A^{k+1}$
\smallskip
\State Obtain $V_{\partial B}^{k+1}$ of $V_A^{k+1}$.
\smallskip
\EndIf
\State $k \leftarrow k + 1$
\EndWhile 

\smallskip

\State $N = k$

\State $k = 0 , \ \nu = 1$

\smallskip

\While{ \textit{not\_converged} }
\Comment{Nonlinear DD loop}

\For{$k < N$} 
\State Collect $\delta u_{\partial B}^{k}$ over $V_{\partial B}^{k}$.
\If{$\left \| \delta u_A^{k, \, \nu} \right \|_{\infty} \geq \epsilon$ $\, \textrm{or} \,$ $\left \| \delta u_{\partial B}^{k} \right \|_{\infty} \geq \epsilon$}

\State Local solve over $V_A^k$,
\smallskip
\State $u^{\nu + 1} = u^{\nu} + R_A^T \, \delta u_A$
\State Check convergence criteria.
\smallskip
\EndIf
\EndFor
\State $\nu \leftarrow \nu + 1$
\EndWhile 

\end{algorithmic}
\end{algorithm}


The new adaptive solver is also outlined in Algorithm \ref{alg:local_nDD}. Note that a subdomain needs to be solved only when the solution is not yet converged or the boundary values change. Therefore the localization is naturally achieved during the nonlinear DD process. This leads to a reliable strategy to exploit the locality for each outer iteration.

An outer iteration of the DD method can be written in a fixed-point form, 
\begin{equation} 
u^{\nu+1} = \sum_{i=1}^{N} R_i^T \, G_i (u^{\nu}) =: \mathcal{G} (u^{\nu})
\end{equation}
where the solution operator for a subproblem is, 
\begin{equation} 
u_{\Omega_i}^{\nu+1} = G_i (u^{\nu})
\end{equation}
As can be seen, evaluation of the function $\mathcal{G} (u)$ involves the solution of all the subproblems $\left ( 1 , ..., N \right )$. Despite its simple form, the fixed-point method may suffer from slow convergence, or even divergence [35,37]. Recently we proposed several ways of accelerating the nonlinear DD process [38]. The nonlinear acceleration techniques greatly improve the outer convergence behavior, while requiring little additional cost. The investigation on the outer convergence of the adaptive DD solver is subject to future work.

\section{Results}



We evaluate the efficacy of the localization methods using several problems with discrete fracture networks. The problems include an oil-water system and a compositional system with phase changes.~A 2D synthetic model is generated to contain a single-stage hydraulically-fractured horizontal well at the center of a reservoir.~The fractures are assumed to fully penetrate the formation.~An embedded discrete fracture model (EDFM) is employed to explicitly describe the discrete fractures. Lee et al. [39], Li and Lee [40], Hajibeygi et al.~[41] and Moinfar et al.~[42] introduced and extended EDFM, which does not require simulation grid to conform to fracture geometry. Recent works on the implementations of EDFM for various types of problems include [8,12-14,43-48].

A simple time-stepping strategy is employed: starting with a small initial value, timestep sizes gradually increase to the maximum value. Newton convergence is based on the following criterion: solution (pressure) delta (increment) $\left \| \delta p \right \|_{\infty} < \epsilon_p$ between iterations. The specification of the base model is given in Table \ref{tab:specification}. 

\begin{table}[!htb]
\centering
\caption{Specification of the synthetic base model.}
\label{tab:specification}
\begin{tabular}{|c|c|c|}
\hline
Parameter                 & Value         & Unit   \\ \hline
Initial pressure          & 2500          & psi    \\ \hline
Matrix porosity           & 0.05          &        \\ \hline
Rock compressibility      & 3.4e-4        & 1/psi  \\ \hline
Matrix permeability       & 1e-19         & $\textrm{m}^2$     \\ \hline
Fracture permeability     & 1e-10         & $\textrm{m}^2$     \\ \hline
Fracture aperture         & 1e-3          & m       \\ \hline
Production BHP            & 1000          & psi     \\ \hline
Total simulation time     & 1500          & day     \\ \hline
Max timestep size         & 100           & day     \\ \hline
\end{tabular}
\end{table}

\subsection{Grid sensitivity}


We first test a $100 \textrm{m} \times 100 \textrm{m}$ model with two fractures to demonstrate the effect of transient flow. Initial water saturation is set as the connate saturation, so that water is immobile during simulations. Newton tolerance has the value of $\epsilon_p = 0.3 \ \textrm{psi}$. Simulations are run for three different levels of grid resolution. 

\begin{figure}[!htb]
\centering
\subfloat[$20 \times 20$]{
\includegraphics[scale=0.33]{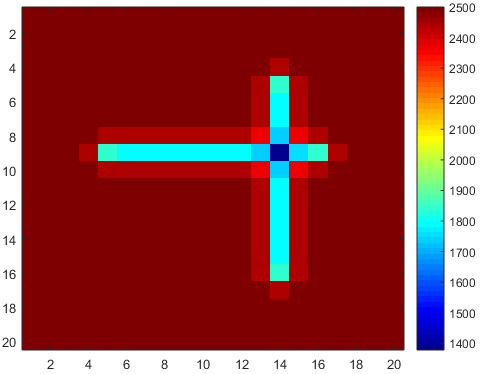}}
\
\subfloat[$50 \times 50$]{
\includegraphics[scale=0.33]{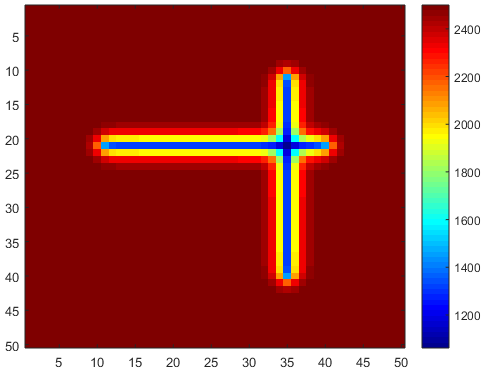}}
\
\subfloat[$200 \times 200$]{
\includegraphics[scale=0.33]{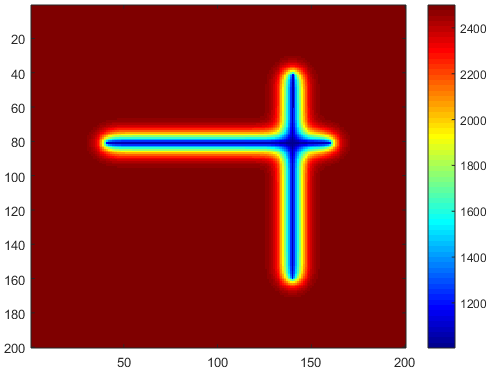}}

\caption{Pressure profiles for the three levels of grid resolution.}
\label{fig:standard_pres}
\end{figure}

\begin{figure}[!htb]
\centering
\includegraphics[scale=0.6]{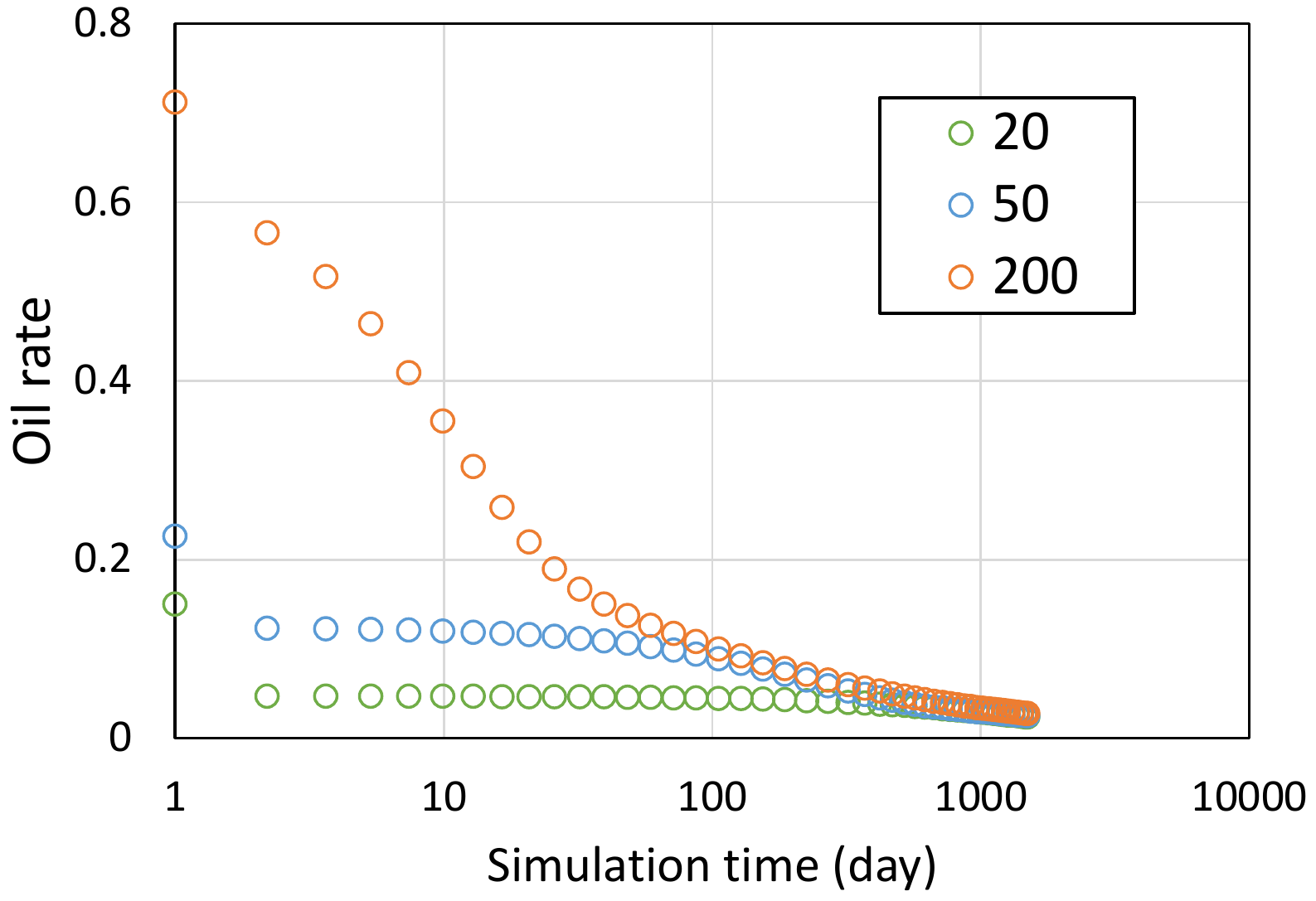}
\caption{Oil rates for the three levels of grid resolution.}
\label{fig:standard_rate}
\end{figure}

Pressure profiles at the end of simulation are shown in \textbf{Fig. \ref{fig:standard_pres}}. Oil rates are plotted in \textbf{Fig.~\ref{fig:standard_rate}}.~As we can see, oil productions are largely underestimated by the coarse grid systems.~The large cell sizes of coarse grid are not adequate for the sharp pressure variations in the vicinity of the fractures. On the other hand, the fine-grid case involves a large number of cells and thus requires significant computational efforts.

\subsection{Localized Newton method}

\subsubsection{Case 1}



\begin{figure}[!htb]
\centering
\includegraphics[scale=0.55]{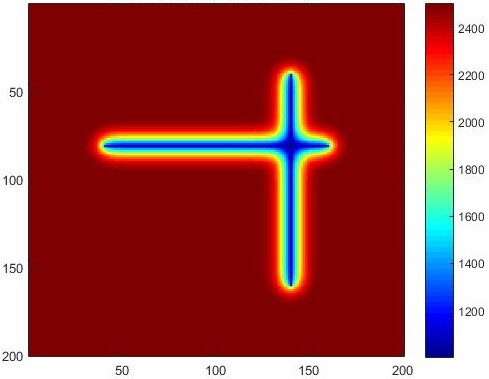}
\caption{Pressure profile of Case 1.}
\label{fig:case_1_pres}
\end{figure}  

\begin{figure}[!htb]
\centering
\includegraphics[scale=0.6]{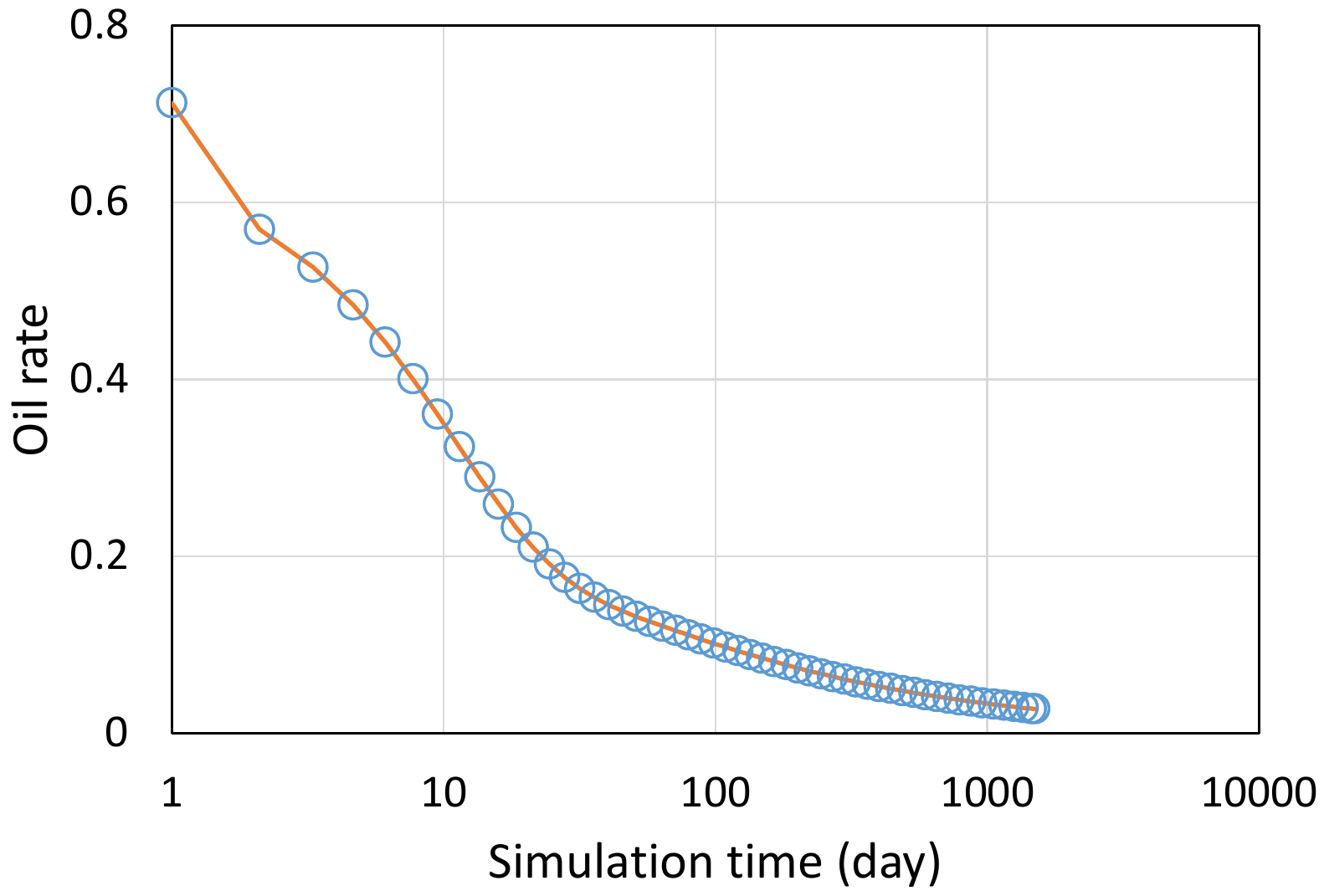}
\caption{Oil rates of Case 1.}
\label{fig:case_1_rate}
\end{figure}  

\begin{figure}[!htb]
\centering
\includegraphics[scale=0.6]{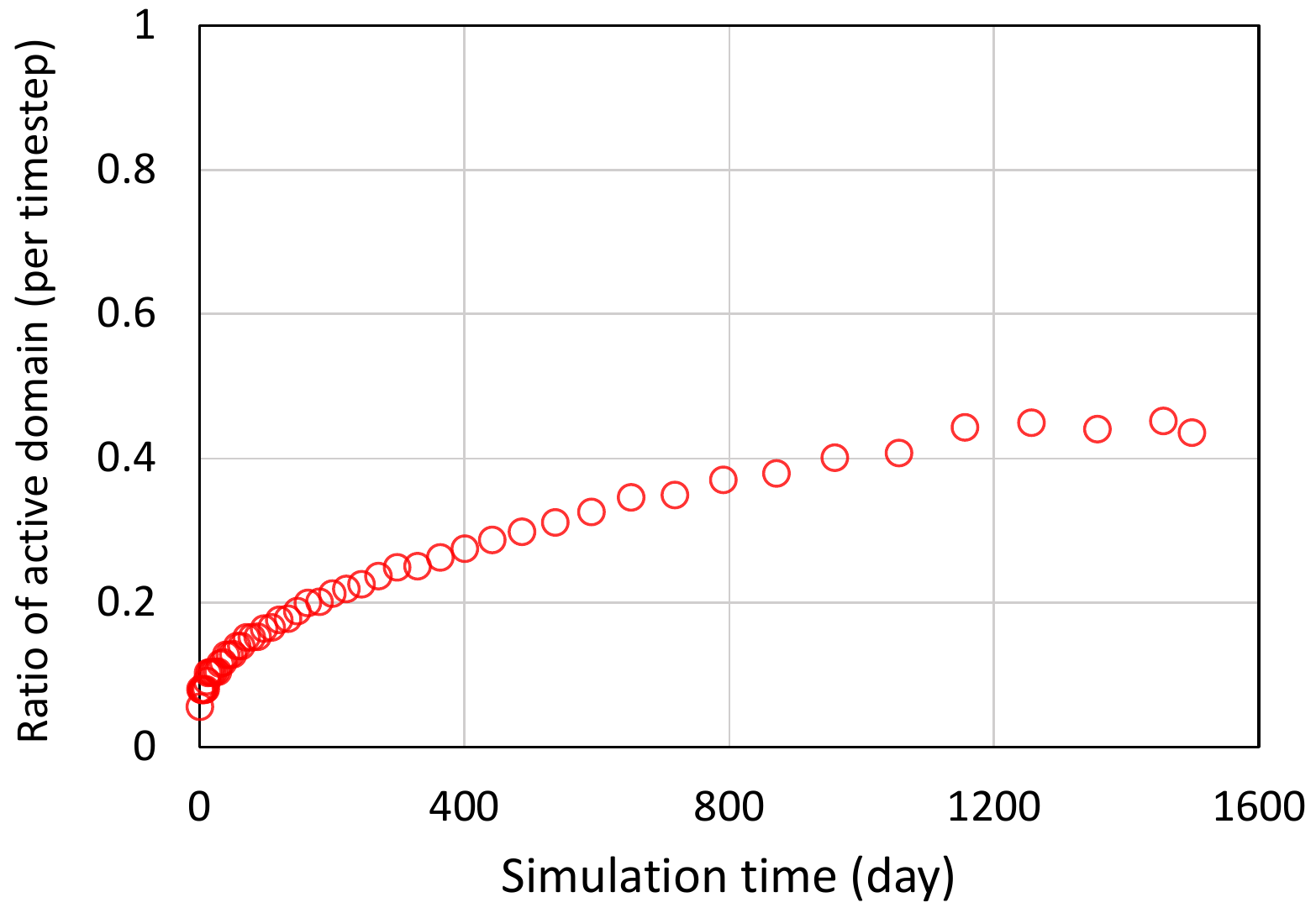}
\caption{Ratios of active domain (per timestep) for Case 1.}
\label{fig:case_1_re}
\end{figure}

We test the model with $200 \times 200$ grid level. At every iteration, the localized Newton algorithm \ref{alg:local_new} provides the active set to be updated, and then solves the reduced linear system. Convergence tolerance of $\epsilon_p = 0.3 \ \textrm{psi}$ is employed as the cutoff value for the active and boundary sets. The neighbor set $V_{nbr}^{i, m}$ with $m=2$ is used to expand active domain. The support set of the last timestep is taken as the initial active set for the current timestep. 

Pressure profile is shown in \textbf{Fig.~\ref{fig:case_1_pres}}.~We can observe that only a small fraction of the cells around the fractures undergoes significant changes of pressure. Oil rates are plotted in \textbf{Fig.~\ref{fig:case_1_rate}}.~As expected, the solution from the localization method exactly matches the reference solution. This is because the iterative processes converge under the same convergence tolerance. 

We plot the ratios of active domain (per timestep) versus simulation time in \textbf{Fig. \ref{fig:case_1_re}}. Note that 2-4 iterations are taken at each timestep. Computational performance of Case 1 is summarized in Table \ref{tab:case_1_tab}. $M_A$ is defined as the ratio of the active to full sets. For the standard Newton, the total ratio $M_A$ is equal to the total iteration number, with the average ratio as 1. 

The results show that the localization method exhibits good convergence performance. As we discussed before, the iterations for expanding active domain can be viewed as nonlinear preconditioning which provides better initial guesses. The localized solver achieves an at least 13-fold reduction in computations, compared to the standard solver. The actual simulation speedup depends on the scaling of computational complexity $O (n^{\beta})$. 

\begin{table}[!htb]
\centering
\caption{Computational performance of Case 1.}
\label{tab:case_1_tab}
\begin{tabular}{|c|c|c|c|c|}
\hline
& Timesteps & \begin{tabular}[c]{@{}c@{}}Total\\ iterations\end{tabular} & {\color[HTML]{3531FF} \begin{tabular}[c]{@{}c@{}} Total ratio \\ $M_A$ \end{tabular}} & \begin{tabular}[c]{@{}c@{}}Average ratio $M_A$ \\ per iteration\end{tabular} \\ \hline
Localized Newton & 54        & 159                                                        & {\color[HTML]{3531FF} 11.4}        & 0.072                      \bigstrut           \\ \hline
Standard Newton  & 54        & 156                                                        & {\color[HTML]{3531FF} 156}         & 1                          \bigstrut           \\ \hline
\end{tabular}
\end{table}

\begin{figure}[!htb]
\centering
\includegraphics[scale=0.4]{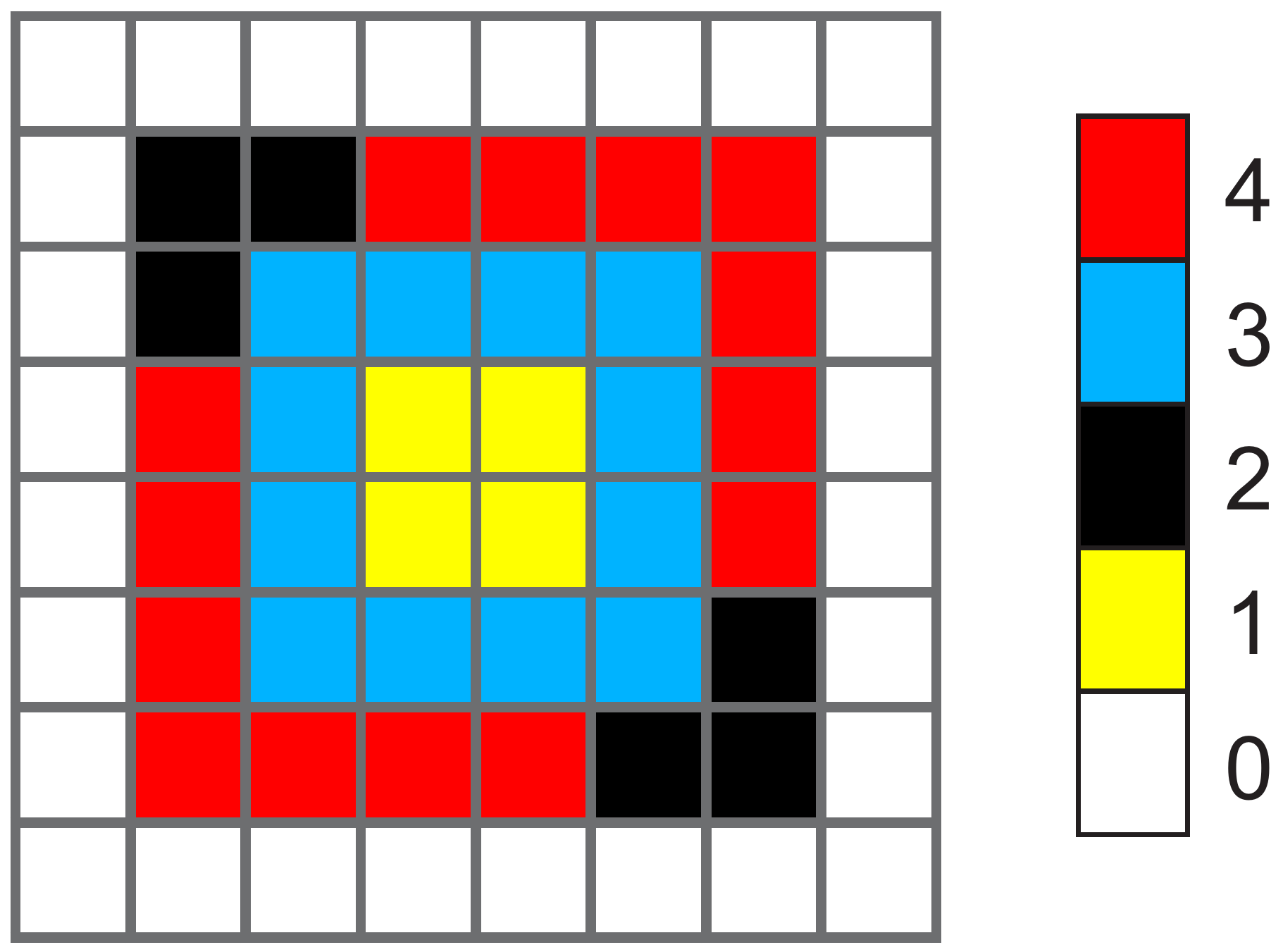}
\caption{Illustration for the cell types: 1. active set (yellow); 2. boundary set (black); 3. the active cells with non-negligible update (blue); 4. the boundary cells with non-negligible update (red).}
\label{fig:case_1_illu}
\end{figure}

\begin{figure}[!htb]
\centering
\subfloat[Iteration 1]{
\includegraphics[scale=0.5]{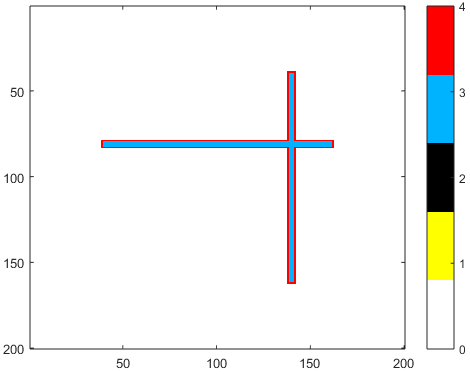}}
\
\subfloat[Iteration 2]{
\includegraphics[scale=0.5]{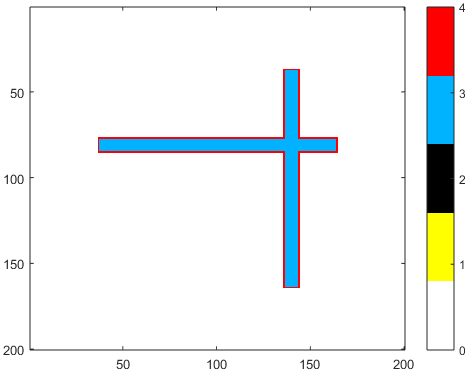}}
\
\subfloat[Iteration 3]{
\includegraphics[scale=0.5]{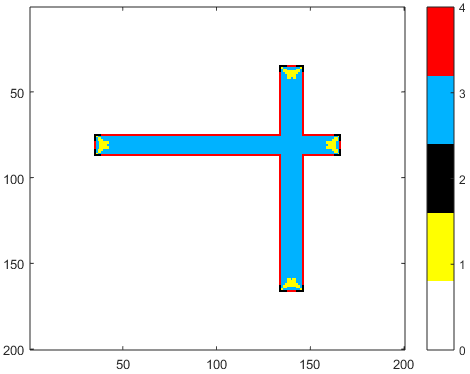}}
\
\subfloat[Iteration 4]{
\includegraphics[scale=0.5]{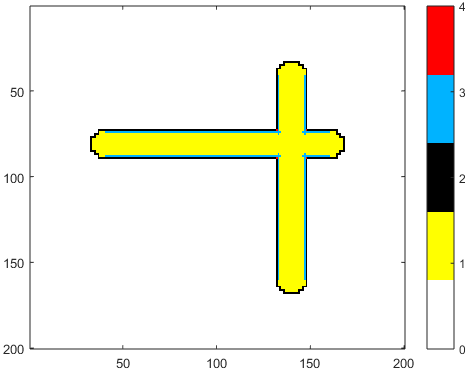}}
\caption{Flagging profiles for Case 1.}
\label{fig:case_1_flag}
\end{figure}

We re-run the case using the localized Newton for one timestep with the size of 50 days. The timestep size is equal to the total simulation time. The profiles for the flagging of cells over the 4 iterations are plotted in \textbf{Fig.~\ref{fig:case_1_flag}}. Four types of cell sets are specified: 1. active set (in color yellow); 2. boundary set (black); 3. the active cells with non-negligible update (the support set), (blue); and 4. the boundary cells with non-negligible update (red). A color illustration for the cell types is given in \textbf{Fig. \ref{fig:case_1_illu}}. 

As can be seen, the initial active set is small and not conservative, resulting in aggressive localization and thus high computational speedup. As the iterations proceed, the active domain expands until the maximum area is reached.

\subsubsection{Case 2}

\begin{figure}[!htb]
\centering
\includegraphics[scale=0.6]{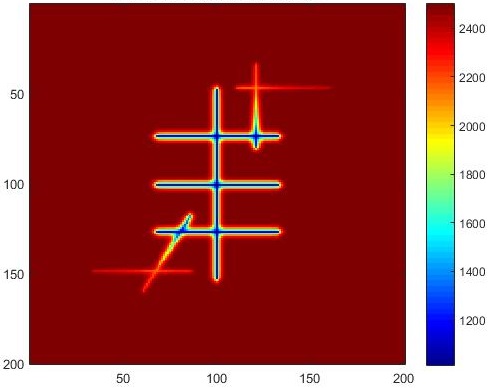}
\caption{Pressure profile of Case 2.}
\label{fig:Case_2_pres}
\end{figure}

\begin{figure}[!htb]
\centering
\includegraphics[scale=0.6]{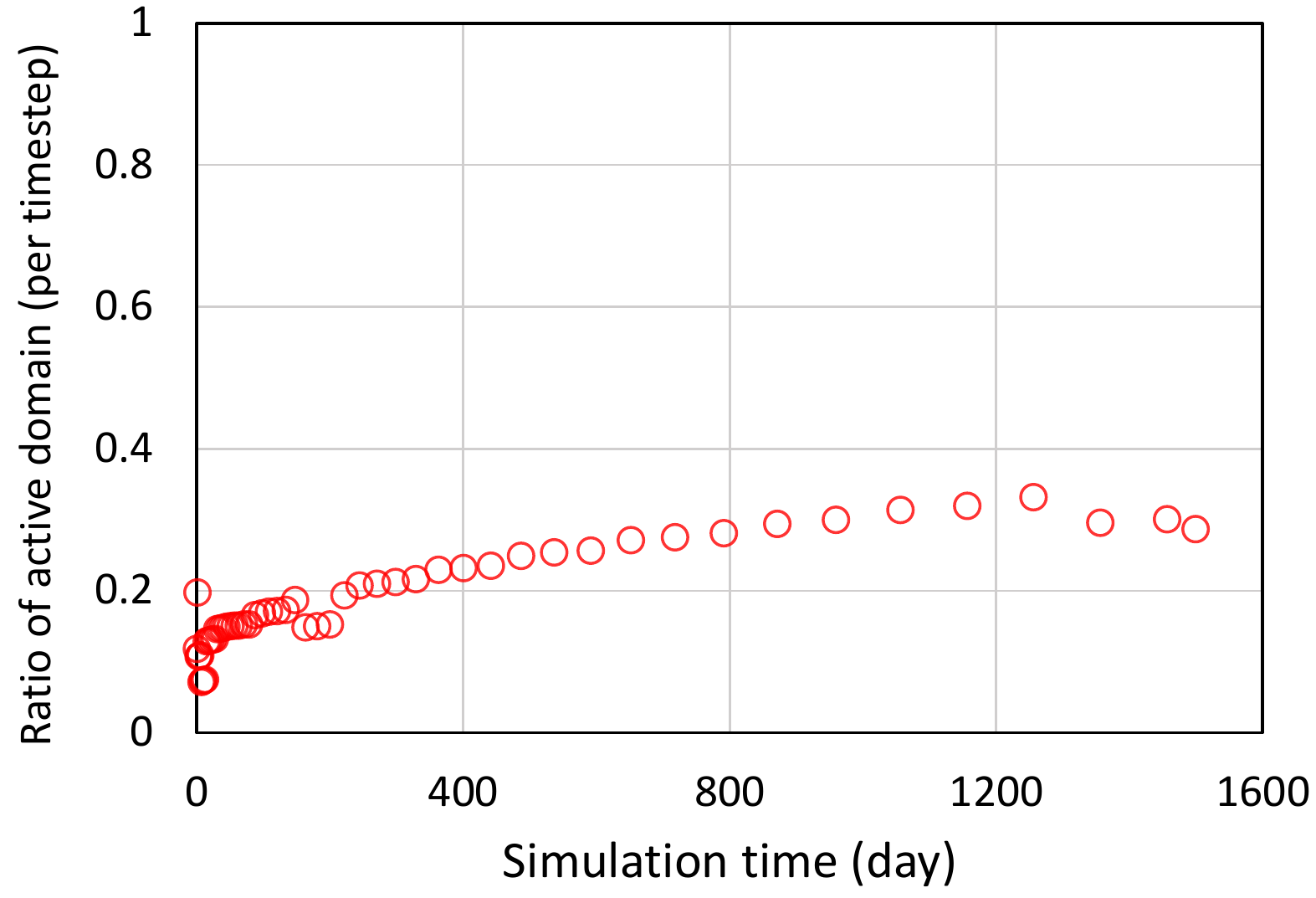}
\caption{Ratios of active domain (per timestep) for Case 2.}
\label{fig:Case_2_re}
\end{figure}

We test a $300 \textrm{m} \times 300 \textrm{m}$ model with $200 \times 200$ grid and a more complex fracture network. The model contains 4 secondary fractures with permeability of 1e-13 $\textrm{m}^2$. Pressure profile is shown in \textbf{Fig.~\ref{fig:Case_2_pres}}. We plot the ratios of active domain (per timestep) versus simulation time in \textbf{Fig.~\ref{fig:Case_2_re}}. 

Computational performance of Case 2 is summarized in Table \ref{tab:Case_2_tab}. From the results we see that the localization method enables a significant reduction in computations, while preserving the original convergence behavior.

\begin{table}[!htb]
\centering
\caption{Computational performance of Case 2.}
\label{tab:Case_2_tab}
\begin{tabular}{|c|c|c|c|c|}
\hline
& Timesteps & \begin{tabular}[c]{@{}c@{}}Total\\ iterations\end{tabular} & {\color[HTML]{3531FF} \begin{tabular}[c]{@{}c@{}} Total ratio \\ $M_A$ \end{tabular}} & \begin{tabular}[c]{@{}c@{}}Average ratio $M_A$ \\ per iteration\end{tabular} \\ \hline
Localized Newton & 54        & 155                                                        & {\color[HTML]{3531FF} 10.1}        & 0.065                      \bigstrut           \\ \hline
Standard Newton  & 54        & 153                                                        & {\color[HTML]{3531FF} 153}         & 1                          \bigstrut           \\ \hline
\end{tabular}
\end{table}

\begin{figure}[!htb]
\centering
\subfloat[Iteration 1]{
\includegraphics[scale=0.5]{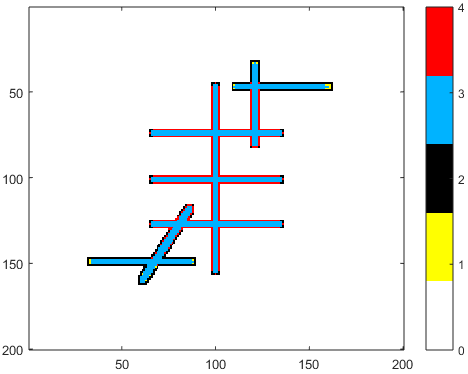}}
\
\subfloat[Iteration 2]{
\includegraphics[scale=0.5]{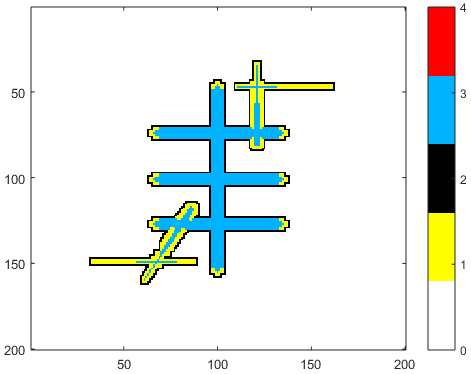}}
\
\subfloat[Iteration 3]{
\includegraphics[scale=0.5]{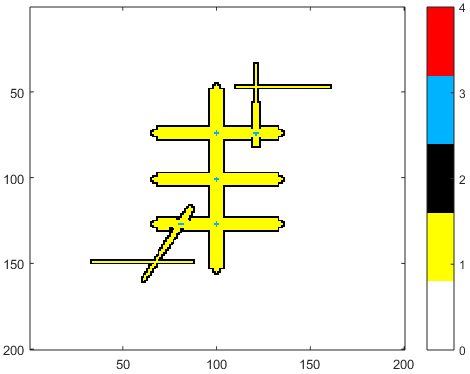}}
\
\subfloat[Iteration 4]{
\includegraphics[scale=0.5]{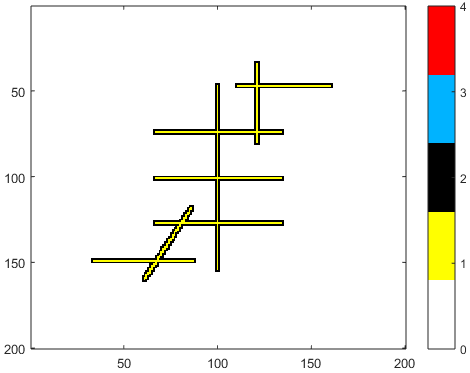}}
\caption{Flagging profiles for Case 2.}
\label{fig:case_2_flag}
\end{figure}

We re-run the case for one timestep with the size of 50 days. The profiles for the flagging of cells over the 4 iterations are plotted in \textbf{Fig.~\ref{fig:case_2_flag}}. The color illustration of cell sets is the same as specified in the previous section. As can be seen, the sparsity patterns of the updates vary largely from one iteration to the next. After two iterations, the nonlinear convergence is constrained to just several cells.

\subsubsection{Case 3}



We again consider the model from Case 2, but with $100 \times 100$ grid. The model uses a two-component fluid system where the initial oil is made of $\left \{ \textrm{C}_1 (50 \%), \textrm{C}_{10} (50 \%) \right \}$. Phase density and viscosity depend on pressure and compositions. Phase molar density $\rho_l$ is evaluated based on the compressibility (Z) factor from the PR EoS. Phase viscosity $\mu_l$ is computed by the correlation of Lohrenz et al. [49]. Simple relative permeabilities given by quadratic function are used. Initial pressure is 2900 psi and temperature is 340 $\textrm{K}$. The total simulation time is 500 days, with the maximum timestep size as 50 days. The other parameters in the previous case remain unchanged.

\begin{figure}[!htb]
\centering
\includegraphics[scale=0.6]{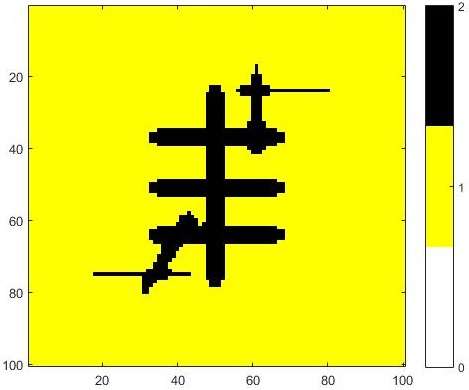}
\caption{Phase status of gas for Case 3.}
\label{fig:case_3_ps}
\end{figure}

\begin{figure}[!htb]
\centering
\includegraphics[scale=0.6]{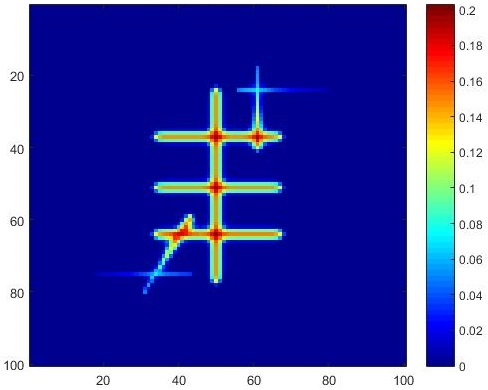}
\caption{Gas saturation for Case 3.}
\label{fig:case_3_Sg}
\end{figure}

The profiles for phase status and gas saturation are shown in \textbf{Fig.~\ref{fig:case_3_ps}} and \textbf{Fig.~\ref{fig:case_3_Sg}}, respectively. As pressure drops below the bubble-point, gas appears around the fractures. The transient effects of saturation and phase dynamics can be difficult to capture using coarse grid. From the results we observe that the saturation updates exhibit considerable locality. By comparison, the pressure updates affect a much larger region.

\begin{figure}[!htb]
\centering
\includegraphics[scale=0.6]{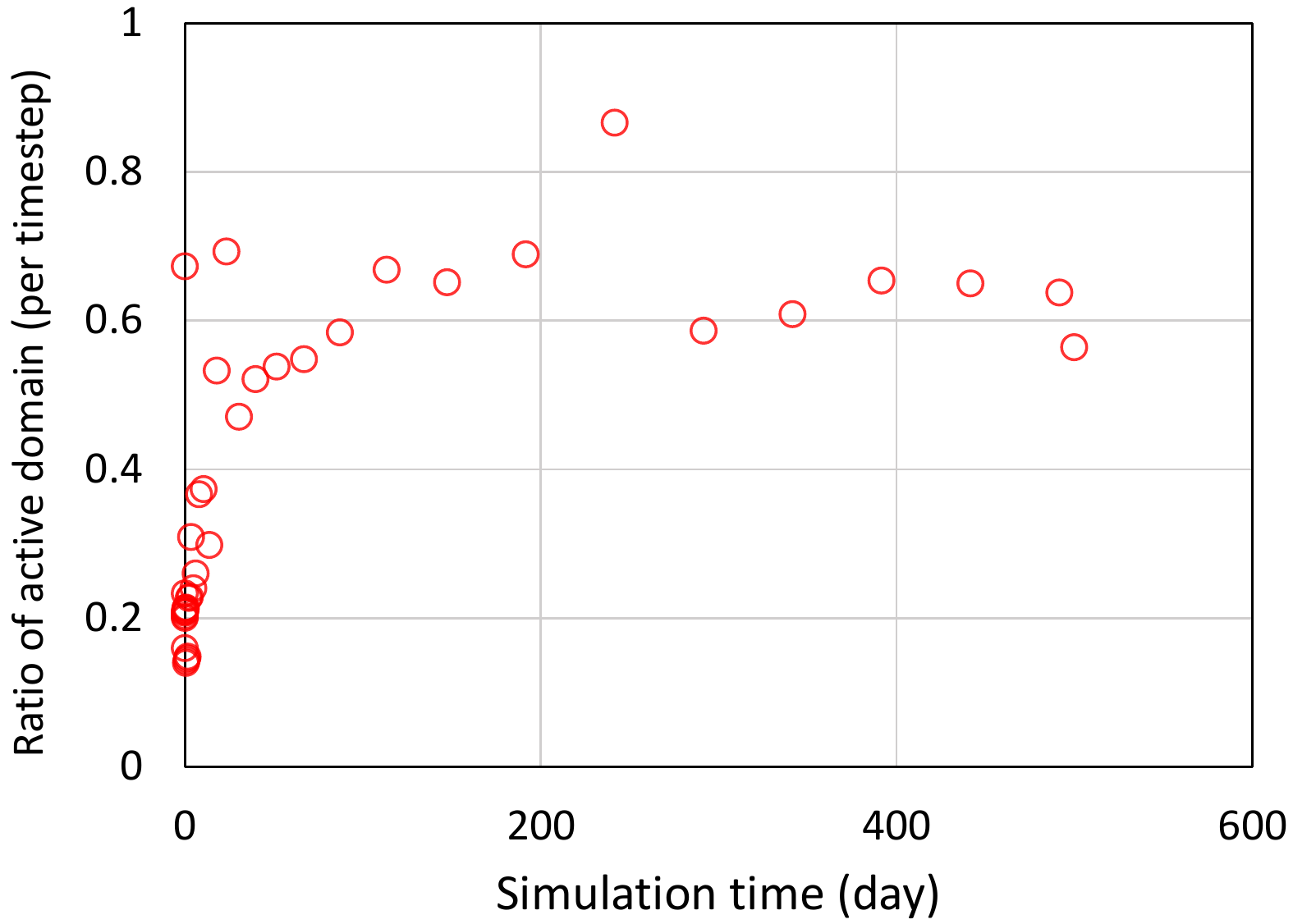}
\caption{Ratios of active domain (per timestep) for Case 3.}
\label{fig:case_3_re}
\end{figure}

We plot the ratios of active domain (per timestep) versus simulation time in \textbf{Fig. \ref{fig:case_3_re}}. Computational performance of Case 3 is summarized in Table \ref{tab:Case_3_tab}. The localized Newton method shows superior performance, with a small increase of iterations.

It should be noted that the ratio $M_A$ may increase with a larger portion of the domain taken by fracture networks.~We do not exploit solution locality for fractures under the current implementations.~Overall computational efforts can be further reduced, if applying localization also to fracture cells. The localized solver will be more effective, for large fracture networks with various conductivities and secondary fractures, because of the transient effects in fracture cells.

\begin{table}[!htb]
\centering
\caption{Computational performance of Case 3.}
\label{tab:Case_3_tab}
\begin{tabular}{|c|c|c|c|c|}
\hline
& Timesteps & \begin{tabular}[c]{@{}c@{}}Total\\ iterations\end{tabular} & {\color[HTML]{3531FF} \begin{tabular}[c]{@{}c@{}} Total ratio \\ $M_A$ \end{tabular}} & \begin{tabular}[c]{@{}c@{}}Average ratio $M_A$ \\ per iteration\end{tabular} \\ \hline
Localized Newton & 40        & 153                                                        & {\color[HTML]{3531FF} 16.1}        & 0.1                        \bigstrut           \\ \hline
Standard Newton  & 40        & 148                                                        & {\color[HTML]{3531FF} 148}         & 1                          \bigstrut           \\ \hline
\end{tabular}
\end{table}

\subsection{Adaptive Nonlinear DD method}

We study the adaptive nonlinear DD (Algorithm \ref{alg:local_nDD}) which is based on dynamic partitions to perform localized computations.~The developed solver can make adequate estimates of the active set for each inner iteration.

\subsubsection{Case 2.1}

\begin{figure}[!htb]
\centering
\includegraphics[scale=0.6]{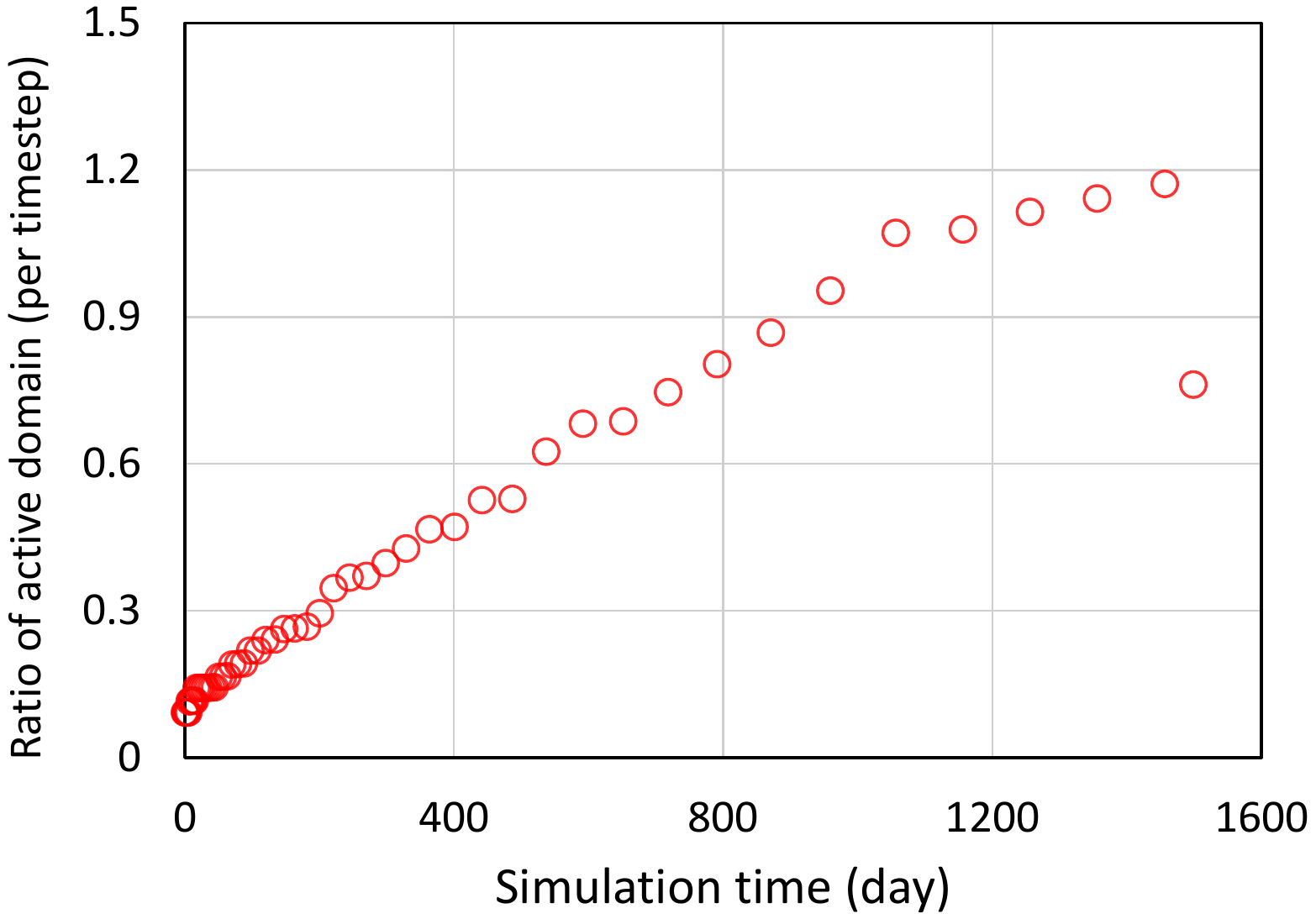}
\caption{Ratios of active domain (per timestep) for Case 2.1.}
\label{fig:case_21_re}
\end{figure}

We use the same model as specified in Case 1. Computational performance of the case is summarized in Table \ref{tab:Case_21_tab}. The adaptive DD solver greatly reduces the computational cost, while taking much more iterations. This is because each outer iteration consists of multiple inner iterations in the nonlinear DD process. We observe that the size of a subdomain system is relatively small and the total number of outer iterations is comparable to the standard Newton method. The ratios of active domain (per timestep) versus simulation time are plotted in \textbf{Fig.~\ref{fig:case_21_re}}.

\begin{table}[!htb]
\centering
\caption{Computational performance of Case 2.1.}
\label{tab:Case_21_tab}
\begin{tabular}{|c|c|c|c|c|}
\hline
& Timesteps & \begin{tabular}[c]{@{}c@{}}Total (inner) \\ iterations\end{tabular} & {\color[HTML]{3531FF} \begin{tabular}[c]{@{}c@{}} Total ratio \\ $M_A$ \end{tabular}} & \begin{tabular}[c]{@{}c@{}}Average ratio $M_A$ \\ per iteration\end{tabular} \\ \hline
Adaptive Nonlinear DD & 54        & 804                                                        & {\color[HTML]{3531FF} 20.8}        & 0.026                      \bigstrut           \\ \hline
Standard Newton  & 54        & 156                                                        & {\color[HTML]{3531FF} 156}         & 1                          \bigstrut           \\ \hline
\end{tabular}
\end{table}

\begin{figure}[!htb]
\centering
\subfloat[Iteration 1]{
\includegraphics[scale=0.5]{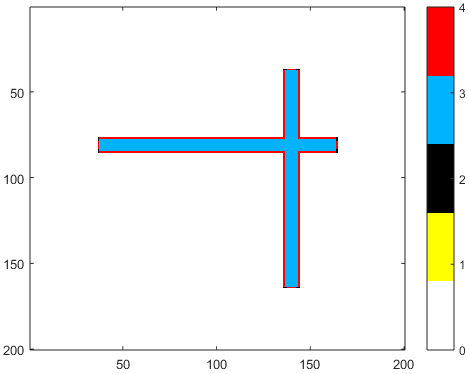}}
\
\subfloat[Iteration 2]{
\includegraphics[scale=0.5]{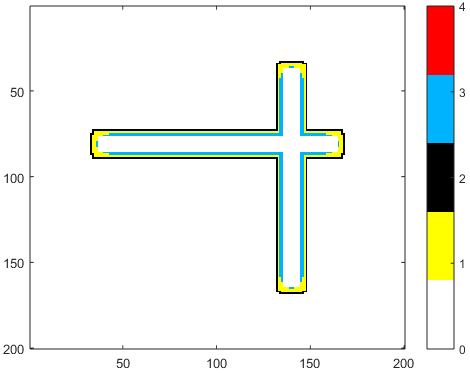}}
\
\subfloat[Iteration 3]{
\includegraphics[scale=0.5]{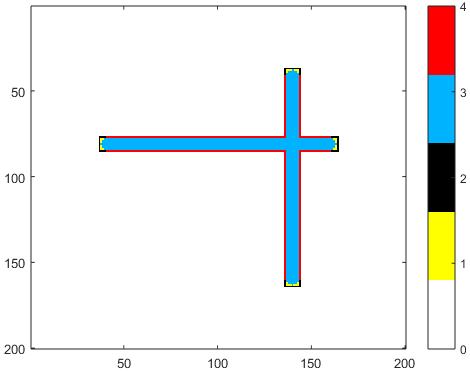}}
\
\subfloat[Iteration 4]{
\includegraphics[scale=0.5]{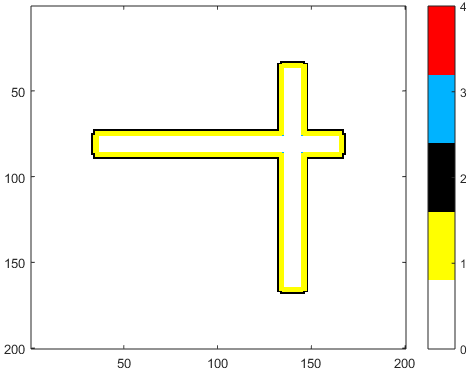}}
\caption{Flagging profiles for Case 2.1.}
\label{fig:case_21_flag}
\end{figure}

We re-run the case for one timestep with the size of 50 days. The neighbor set $V_{nbr}^{i, m}$ is specified with $m=4$. Flagging profiles of cells over the 4 iterations are plotted in \textbf{Fig.~\ref{fig:case_21_flag}}.~As we can see, the timestep converges after 2 outer iterations. The algorithm constructs the subdomains that adapt to the flow dynamics and pressure updates. During the nonlinear DD process, the subproblems are solved sequentially, to achieve localization. The results confirm that the support set of the timestep is contained in the union of all the flagged subsets.

\subsubsection{Case 2.2}

\begin{figure}[!htb]
\centering
\includegraphics[scale=0.6]{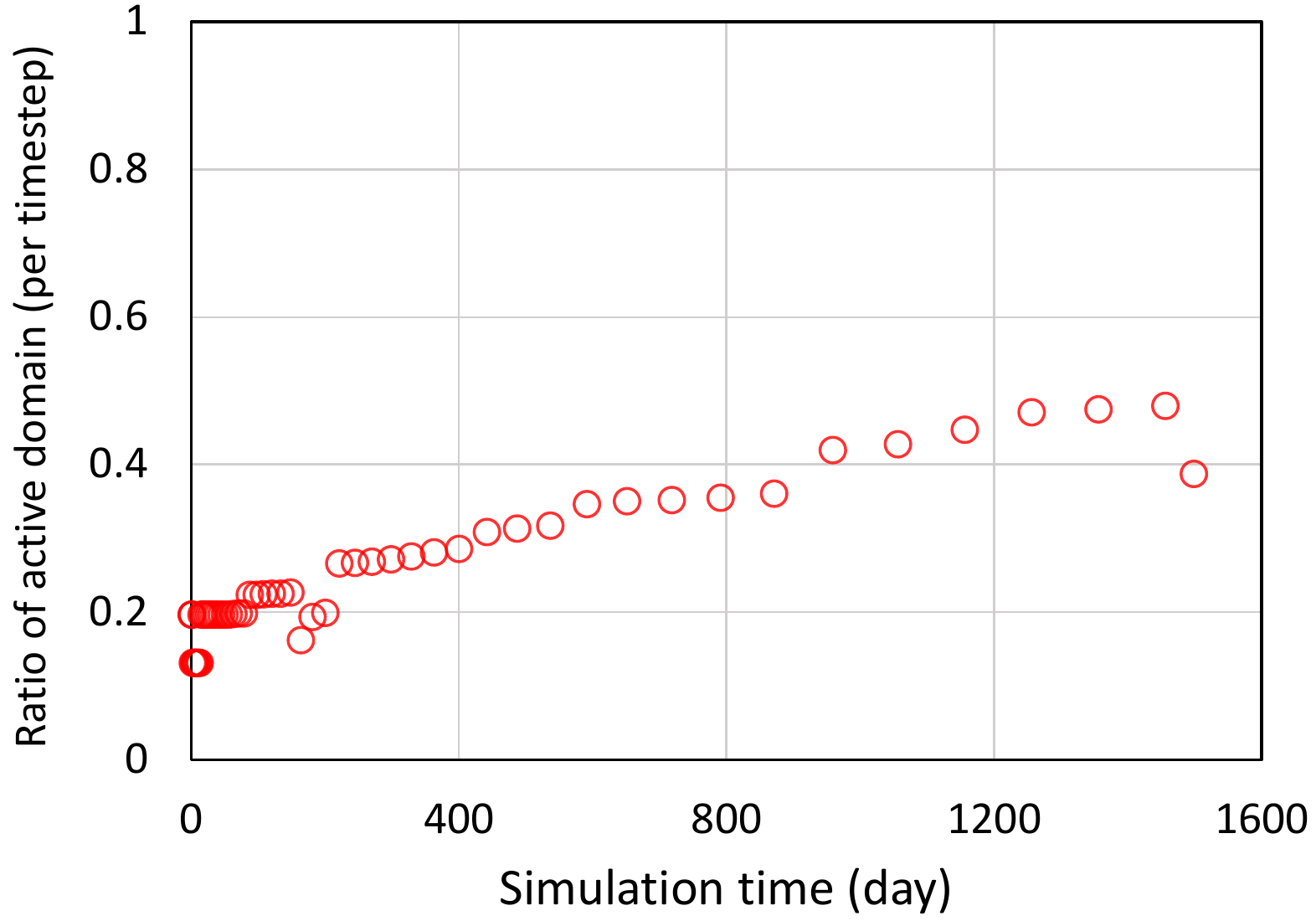}
\caption{Ratios of active domain (per timestep) for Case 2.2.}
\label{fig:case_22_re}
\end{figure}

The model from Case 2 is used. Computational performance of the case is summarized in Table \ref{tab:Case_22_tab}. We plot the ratios of active domain (per timestep) versus simulation time in \textbf{Fig.~\ref{fig:case_22_re}}. We can see that the adaptive solver obtains an at least 11-fold reduction in solution effort, without degrading the original Newton convergence.

\begin{table}[!htb]
\centering
\caption{Computational performance of Case 2.2.}
\label{tab:Case_22_tab}
\begin{tabular}{|c|c|c|c|c|}
\hline
& Timesteps & \begin{tabular}[c]{@{}c@{}}Total (inner) \\ iterations\end{tabular} & {\color[HTML]{3531FF} \begin{tabular}[c]{@{}c@{}} Total ratio \\ $M_A$ \end{tabular}} & \begin{tabular}[c]{@{}c@{}}Average ratio $M_A$ \\ per iteration\end{tabular} \\ \hline
Adaptive Nonlinear DD & 54        & 255                                                        & {\color[HTML]{3531FF} 13.6}        & 0.053                      \bigstrut           \\ \hline
Standard Newton  & 54        & 153                                                        & {\color[HTML]{3531FF} 153}         & 1                          \bigstrut           \\ \hline
\end{tabular}
\end{table}

This work develops a prototype algorithm for the adaptive DD solver, which can be further improved and optimized, e.g.~exploiting the solution locality within each subdomain. The algorithm will be expected to prevent any overly conservative estimate. Subsequently, higher simulation speedup can be obtained.

We note that solution strategies of dynamic-type introduce additional challenges associated with memory management and parallelization. Practical implementations of the localized algorithms for achieving optimal efficiency are the subject of future work.

\section{Summary}

We develop the localized solution strategies for efficient simulations of unconventional reservoirs. By utilizing the diffusive nature of pressure updates, an adaptive algorithm is proposed to make adequate estimates for the active sets to be solved. We also develop a localized solver based on nonlinear domain decomposition (DD). The solver provides effective partitioning that adapts to flow dynamics and Newton updates. 

We test several complex problems with discrete fracture networks. The results show that large degrees of solution locality present across timesteps and iterations. Compared to a standard Newton solver, the developed solvers enable superior computational performance. Moreover, the original Newton convergence is preserved, without any impact on the solution accuracy. 

The new solvers can be extended to account for more complex fracture networks and flow physics. The incorporation of models with strong heterogeneity and field-scale fracture networks is a subject of our ongoing research.

\section*{Acknowledgements}

This work was supported by the Chevron/Schlumberger INTERSECT Research \& Prototyping project. The author thanks Chevron for permission to publish this paper. The author also thanks Rami Younis at University of Tulsa for constructive discussions.

\section*{Appendix A. Discretization methods}

A standard finite-volume scheme is applied as the spatial discretization for the mass conservation equations. A two-point flux approximation (TFPA) is used to approximate the flux across a cell interface. The method of choice for the time discretization is the fully-implicit scheme. The discrete form of conservation equation is given as, 
\begin{equation} 
\frac{V}{\Delta t} \left [ \left ( \phi \rho_T z_c \right )^{n+1} - \left ( \phi \rho_T z_c \right )^{n} \right ] - \sum_{ij}\left ( x_{c} \rho_o F_o + y_{c} \rho_g F_g \right )^{n+1} - Q_c^{n+1} = 0.
\end{equation}
where superscripts denote timesteps, and $\Delta t$ is the timestep size. $V$ is the cell volume. All indices related to the cell numeration are neglected. The accumulation term involves the total density, 
\begin{equation} 
\rho_T z_c = x_{c} \rho_o s_o + y_{c} \rho_g s_g 
\end{equation}
and, 
\begin{equation} 
\rho_T = \begin{cases}
s_o \rho_o (\textbf{x}) + s_g \rho_g (\textbf{y}) , \ \ \textrm{two phase}, \\ 
\rho_l (\textbf{z}) , \ \ \textrm{one phase}.
\end{cases}
\end{equation}
where $\rho_o (\textbf{x})$ indicates the density computed at a composition $\textbf{x}$, and $\rho_l (\textbf{z})$ is the density computed in the single-phase regime at a composition $\textbf{z}$.

The discrete phase flux across the interface $(ij)$ between two cells is written as, 
\begin{equation} 
F_{l,ij} = \Upsilon_{ij} \lambda_{l,ij} \Delta \Phi_{l,ij}
\end{equation}
where subscript $(ij)$ denotes quantities defined at the cell interface. $\Upsilon_{ij}$ is the interface transmissibility. $\Delta \Phi_{l,ij} = \Delta p_{ij} - g_{l,ij}$ is the phase potential difference with the discrete weights $g_{l,ij} = \rho_{l,ij} \ g \Delta h_{ij}$. The phase and compositional coefficients associated with the flux terms are evaluated using the Phase-Potential Upwinding (PPU) scheme.

\section*{References}

[1] S.C. Maxwell, T.I. Urbancic, N. Steinsberger, R. Zinno, others, Microseismic imaging of hydraulic fracture complexity in the Barnett shale, in: SPE Annu. Tech. Conf. Exhib., 2002.

[2] M.K. Fisher, C.A. Wright, B.M. Davidson, A.K. Goodwin, E.O. Fielder, W.S. Buckler, N.P. Steinsberger, others, Integrating fracture mapping technologies to optimize stimulations in the Barnett Shale, in: SPE Annu. Tech. Conf. Exhib., 2002.

[3] M.J. Mayerhofer, E. Lolon, N.R. Warpinski, C.L. Cipolla, D.W. Walser, C.M. Rightmire, others, What is stimulated reservoir volume?, SPE Prod. Oper. 25 (2010) 89–98.

[4] C.L. Cipolla, E.P. Lolon, J.C. Erdle, B. Rubin, others, Reservoir modeling in shale-gas reservoirs, SPE Reserv. Eval. Eng. 13 (2010) 638–653.

[5] X. Weng, O. Kresse, D. Chuprakov, C.-E. Cohen, R. Prioul, U. Ganguly, Applying complex fracture model and integrated workflow in unconventional reservoirs, J. Pet. Sci. Eng. 124 (2014) 468–483.

[6] C.L. Cipolla, N.R. Warpinski, M. Mayerhofer, E.P. Lolon, M. Vincent, others, The relationship between fracture complexity, reservoir properties, and fracture-treatment design, SPE Prod. Oper. 25 (2010) 438–452.

[7] D.Y. Ding, Y.S. Wu, L. Jeannin, Efficient simulation of hydraulic fractured wells in unconventional reservoirs, J. Pet. Sci. Eng. 122 (2014) 631–642.

[8] P. Panfili, R. Colin, A. Cominelli, D. Giamminonni, L. Guerra, others, Efficient and effective field scale simulation of hydraulic fractured wells: methodology and application, in: SPE Reserv. Characterisation Simul. Conf. Exhib., 2015.

[9] M. Karimi-Fard, L.J. Durlofsky, K. Aziz, others, An efficient discrete fracture model applicable for general purpose reservoir simulators, in: SPE Reserv. Simul. Symp., 2003.

[10] B.T. Mallison, M.-H. Hui, W. Narr, Practical gridding algorithms for discrete fracture modeling workflows, in: ECMOR XII-12th Eur. Conf. Math. Oil Recover., 2010: p. cp--163.

[11] V. Artus, D. Fructus, Transmissibility corrections and grid control for shale gas numerical production forecasts, Oil Gas Sci. Technol. d’IFP Energies Nouv. 67 (2012) 805–821.

[12] J. Jiang, R.M. Younis, Hybrid coupled discrete-fracture/matrix and multicontinuum models for unconventional-reservoir simulation, SPE J. 21 (2016).

[13] M. HosseiniMehr, M. Cusini, C. Vuik, H. Hajibeygi, Algebraic dynamic multilevel method for embedded discrete fracture model (F-ADM), J. Comput. Phys. 373 (2018) 324–345.

[14] X. Xue, A. Rey, P. Muron, G. Dufour, X.-H. Wen, others, Simplification and Simulation of Fracture Network Using Fast Marching Method and Spectral Clustering for Embedded Discrete Fracture Model, in: SPE Hydraul. Fract. Technol. Conf. Exhib., 2019.

[15] K. Pruess, T.N. Narasimhan, A practical method for modeling fluid and heat flow in fractured porous media, Soc Petrol Eng J. (1982).

[16] J.E. Warren, P.J. Root, others, The behavior of naturally fractured reservoirs, Soc. Pet. Eng. J. 3 (1963) 245–255.

[17] H. Kazemi, L.S. Merrill Jr, K.L. Porterfield, P.R. Zeman, others, Numerical simulation of water-oil flow in naturally fractured reservoirs, Soc. Pet. Eng. J. 16 (1976) 317–326.

[18] M.-H. Hui, B.T. Mallison, M.H. Fyrozjaee, W. Narr, others, The upscaling of discrete fracture models for faster, coarse-scale simulations of IOR and EOR processes for fractured reservoirs, in: SPE Annu. Tech. Conf. Exhib., 2013.

[19] Y.-S. Wu, J. Li, D. Ding, C. Wang, Y. Di, others, A generalized framework model for the simulation of gas production in unconventional gas reservoirs, Spe J. 19 (2014) 845–857.

[20] J. Jiang, R.M. Younis, A multimechanistic multicontinuum model for simulating shale gas reservoir with complex fractured system, Fuel. 161 (2015).

[21] N. Farah, D.Y. Ding, Y.S. Wu, others, Simulation of the impact of fracturing fluid induced formation damage in shale gas reservoirs, in: SPE Reserv. Simul. Symp., 2015.

[22] G.T. Ren, J.M. Jiang, R.M. Younis, XFEM-EDFM-MINC for coupled geomechanics and flow in fractured reservoirs, in: 15th Eur. Conf. Math. Oil Recover. ECMOR 2016, 2016.

[23] D.Y. Ding, N. Farah, B. Bourbiaux, Y.-S. Wu, I. Mestiri, others, Simulation of matrix/fracture interaction in low-permeability fractured unconventional reservoirs, SPE J. 23 (2018) 1–389.

[24] K. Aziz and A. Settari. Petroleum reservoir simulation. Chapman \& Hall, 1979.

[25] K. Coats, An equation of state compositional model, SPE J. 20(05) (1980) 363–376.

[26] R. Younis, H.A. Tchelepi, K. Aziz, others, Adaptively Localized Continuation-Newton Method--Nonlinear Solvers That Converge All the Time, SPE J. 15 (2010) 526–544.

[27] M. Cusini, C. van Kruijsdijk, H. Hajibeygi, Algebraic dynamic multilevel (ADM) method for fully implicit simulations of multiphase flow in porous media, J. Comput. Phys. 314 (2016) 60–79.

[28] M. Cusini, B. Fryer, C. van Kruijsdijk, H. Hajibeygi, Algebraic dynamic multilevel method for compositional flow in heterogeneous porous media, J. Comput. Phys. 354 (2018) 593–612.

[29] S.M. Sheth, R.M. Younis, others, Localized linear systems in sequential implicit simulation of two-phase flow and transport, SPE J. 22 (2017) 1–542.

[30] P. Lu, B.L. Beckner, others, An adaptive Newton’s method for reservoir simulation, in: SPE Reserv. Simul. Symp., 2011.

[31] J. Jiang, H.A. Tchelepi, Dissipation-based continuation method for multiphase flow in heterogeneous porous media, J. Comput. Phys. 375 (2018) 307–336.

[32] D. V Voskov, H.A. Tchelepi, Comparison of nonlinear formulations for two-phase multi-component EoS based simulation, J. Pet. Sci. Eng. 82 (2012) 101–111.

[33] M.L. Michelsen, The isothermal flash problem. Part I. Stability, Fluid Phase Equilib. 9 (1982) 1–19.

[34] M.L. Michelsen, The isothermal flash problem. Part II. Phase-split calculation, Fluid Phase Equilib. 9 (1982) 21–40.

[35] V. Dolean, M.J. Gander, W. Kheriji, F. Kwok, R. Masson, Nonlinear preconditioning: How to use a nonlinear Schwarz method to precondition Newton’s method, SIAM J. Sci. Comput. 38 (2016) A3357--A3380.

[36] X.-C. Cai, W.D. Gropp, D.E. Keyes, R.G. Melvin, D.P. Young, Parallel Newton--Krylov--Schwarz algorithms for the transonic full potential equation, SIAM J. Sci. Comput. 19 (1998) 246–265.

[37] J.O. Skogestad, E. Keilegavlen, J.M. Nordbotten, Domain decomposition strategies for nonlinear flow problems in porous media, J. Comput. Phys. 234 (2013) 439–451.

[38] J. Jiang, H.A. Tchelepi, Nonlinear acceleration of sequential fully implicit (SFI) method for coupled flow and transport in porous media, Comput. Methods Appl. Mech. Eng. 352 (2019) 246–275.

[39] S.H. Lee, M.F. Lough, C.L. Jensen, Hierarchical modeling of flow in naturally fractured formations with multiple length scales, Water Resour. Res. 37 (2001) 443–455.

[40] L. Li, S.H. Lee, others, Efficient field-scale simulation of black oil in a naturally fractured reservoir through discrete fracture networks and homogenized media, SPE Reserv. Eval. Eng. 11 (2008) 750–758.

[41] H. Hajibeygi, D. Karvounis, P. Jenny, A hierarchical fracture model for the iterative multiscale finite volume method, J. Comput. Phys. 230 (2011) 8729–8743.

[42] A. Moinfar, A. Varavei, K. Sepehrnoori, R.T., Johns. Development of an Efficient Embedded Discrete Fracture Model for 3D Compositional Reservoir Simulation in Fractured Reservoirs, SPE Journal, 19(02), (2014) pp.289-303.

[43] J. Jiang, R.M. Younis, An improved projection-based embedded discrete fracture model (pEDFM) for multiphase flow in fractured reservoirs, Adv. Water Resour. 109 (2017).

[44] M. Ţene, S.B.M. Bosma, M.S. Al Kobaisi, H. Hajibeygi, Projection-based embedded discrete fracture model (pEDFM), Adv. Water Resour. 105 (2017) 205–216.

[45] G. Ren, J. Jiang, R.M. Younis, A Model for coupled geomechanics and multiphase flow in fractured porous media using embedded meshes, Adv. Water Resour. 122 (2018) 113–130.

[46] M.-H. Hui, G. Dufour, S. Vitel, P. Muron, R. Tavakoli, M. Rousset, A. Rey, B. Mallison, others, A Robust Embedded Discrete Fracture Modeling Workflow for Simulating Complex Processes in Field-Scale Fractured Reservoirs, in: SPE Reserv. Simul. Conf., 2019.

[47] X. Xue, C. Yang, T. Onishi, M.J. King, A. Datta-Gupta, others, Modeling Hydraulically Fractured Shale Wells Using the Fast-Marching Method With Local Grid Refinements and an Embedded Discrete Fracture Model, SPE J. (2019).

[48] A. Rey, J. Schembre, X.-H. Wen, others, Calibration of the Water Flowback in Unconventional Reservoirs with Complex Fractures using Embedded Discrete Fracture Model EDFM, in: SPE Liq. Basins Conf. Am., 2019.

[49] J. Lohrenz, B.G. Bray, C.R. Clark, others, Calculating viscosities of reservoir fluids from their compositions, J. Pet. Technol. 16 (1964) 1–171.

\end{document}